\documentclass[aps,plb,preprintnumbers,nofootinbib]{revtex4}

\usepackage{amsmath,amsfonts,amssymb,bm}
\usepackage{graphicx}
\usepackage{color}
\usepackage{subfigure}
\usepackage{multirow}
\usepackage{textcomp}
\usepackage{slashed}
\usepackage{enumitem}

\definecolor{purple}{rgb}{0.5,0,0.5}
\definecolor{blue}{rgb}{0.0,0,0.9}

\usepackage[colorlinks=true, pdfstartview=FitV, linkcolor=purple, citecolor= purple, urlcolor=blue]{hyperref}

\begin{document}

\title{\Large Probing $\chi _{cJ}(J=0,1,2)$ decay into baryon and  anti-baryon \\with SU(3) flavor analysis}
\author{Bo Lan, Xiu-Ping Fan, and Ru-Min Wang$^\dag$\\
{\scriptsize College of Physics and Communication Electronics, Jiangxi Normal University, Nanchang, Jiangxi 330022, China}\\
$^\dag${\scriptsize Corresponding author.~~Email:ruminwang@sina.com}
 }

\begin{abstract}
With the accurate measurements of $\chi _{cJ}(J=0,1,2)$ charmonium  decays,   we explore  $\chi _{cJ}\to \mathcal{B}_{8}\bar{\mathcal{B}}_{8}$ and $\mathcal{B}_{10}\bar{\mathcal{B}}_{10}$  decays
based on the SU(3) flavor symmetry model, where $\mathcal{B}_{8}$ and $\mathcal{B}_{10}$ are light  octet and decuplet baryons, respectively. The decay amplitude relations are determined by an effective interaction Hamiltonian.
Then using   experimental data and the amplitude relations of $\chi _{cJ}\to \mathcal{B}_{8}\bar{\mathcal{B}}_{8}$, $\mathcal{B}_{10}\bar{\mathcal{B}}_{10}$ decays, relative nonperturbative coefficients
are constrained, and the branching ratios of unmeasured $\chi _{cJ}\to \mathcal{B}_{8}\bar{\mathcal{B}}_{8},\mathcal{B}_{10}\bar{\mathcal{B}}_{10}$, for examples,  $\chi _{cJ}\to n\bar{n}$ and $\chi _{cJ}\to\Lambda \bar{\Sigma}^{0}+\bar{\Lambda} \Sigma^{0}$ channels, are predicted.
Moreover, we discuss the case of adding a mixing angle between $\Lambda $ and $\Sigma ^{0}$, which is determined by the quark mass differences.
Our results  provide valuable insights that will aid in advancing our understanding of the mechanisms and characteristics of $\chi _{cJ}$ decays.

\end{abstract}

\maketitle

\section{Introduction}
Since the discovery of $J/\psi$ meson at SLAC  and BNL experiments \cite{E598:1974sol,SLAC-SP-017:1974ind}, research on charmonium $(c\bar{c})$ has gradually developed \cite{Eichten:1974af,E288:1977xhf,Eichten:1978tg}.
Understanding charmonium decays into baryon and anti-baryon pairs could provide insights into various mechanisms, such as the interference between the strong and electromagnetic interactions, contributions from color octet and singlet states, the pQCD $12\%$ rule in $J/\psi$ and $\psi'$ decays,  SU(3) flavor symmetry breaking effects, etc.
The $\chi_{cJ}$ mesons, being P-wave $c\bar{c}$ triple-states with a spin parity $J^{++}$, have been subject to study in the context of their decay into baryon pairs, utilizing frameworks such as perturbative QCD framework \cite{Wong:1999hc,Liu:2010um,Brodsky:1981kj} and quark-pair creation model \cite{Ping:2004sh}.
Some  $\chi_{cJ}$ meson decays into baryon  pairs, such as $\chi_{c1,2}\to p\overline{p}$,   have long
presented a theoretical challenge  \cite{Wong:1999hc}.  Further studies of the $\chi_{cJ}$ decays could enrich our knowledge of the nature of these charmonium states.

With the development of heavy quark effective theory (HQET) \cite{Isgur:1989vq,Neubert:1993mb,Bigi:1992su}, researchers have acquired
effective approaches to depict the physics of hadronic decays that involve heavy quarks.
However, the physics
of decays involving charmed hadrons presents a challenge for traditional QCD-based methods, such as QCD factorization \cite{Beneke:1999br,Collins:1989gx,Beneke:2001ev}, perturbative QCD \cite%
{Georgi:1974wnj,Dokshitzer:1977sg}, and soft collinear effective theory \cite{Bauer:2000ew,Becher:2014oda,Bauer:2000yr}, since the mass of the charm quark is not heavy enough to undergo significant heavy-quark expansion but not light enough for perturbation theory to be effective \cite{Cheng:2010ry,Cheng:2012wr}.
In the lack of reliable calculations, the symmetry analysis can provide very useful information about the decays, such as SU(3) flavor
symmetry \cite{Gell-Mann:1961omu,Gell-Mann:1962yej}.  The SU(3) flavor symmetry approach, which is independent of the detailed dynamics,
offers an opportunity to relate different decay modes.

Based on the SU(3) flavor symmetry methods, significant progress has been
made in $J/\psi$ and  $\Psi(2S)$ decays \cite{Wei:2009zzh,Mo:2024jjd,Mo:2023wrf,Mo:2021asa,Kopke:1988cs,BaldiniFerroli:2019abd,Ferroli:2020mra}.
However, there is relatively little attention given to the $\chi _{cJ}$ mesons \cite{Ping:2004wm,Zhou:2004mw}, which also belong to the charmonium family.
In this work, we will study  $\chi_{cJ}$ mesons decay into a baryon-antibaryon pair via strong and electromagnetic interactions based on the SU(3) flavor symmetry method.
We will firstly derive the amplitude relations for the $\chi_{cJ}\to \mathcal{B}\bar{\mathcal{B}}$  decays taking into account the research methods used for $J/\psi $ and $\Psi (2S)$ in Refs. \cite{Mo:2023wrf,Ferroli:2020mra},
and then constrain relative nonperturbative parameters and  obtain the branching ratio results for all $\chi _{cJ}\to \mathcal{B}_8\bar{\mathcal{B}}_8,\mathcal{B}_{10}\bar{\mathcal{B}}_{10}$. Furthermore, we will
analyze the case of adding a mixing between $\Lambda $ and $\Sigma ^{0}$.

This paper is organized as follows: In Sec. II,  the amplitude relations  and predicted branching ratios of $\chi _{cJ}\rightarrow \mathcal{B}_8\bar{\mathcal{B}}_8$
are given under three different cases. In Sec. III, the amplitude relations  and predicted branching ratios of $\chi_{cJ}\rightarrow \mathcal{B}_{10}\bar{\mathcal{B}}_{10}$ are given. Our conclusion
is presented in Sec. IV.

\section{Decays  $\chi _{cJ}\to \mathcal{B}_8\bar{\mathcal{B}}_8$}

\subsection{Amplitude Relations}
The light baryon octet  under the SU(3) flavor symmetry of $u,d,s$ quarks
can be written as
\begin{equation}
\mathcal{B}_8=\left(
\begin{array}{ccc}
\frac{\Lambda}{\sqrt{6}}+\frac{\Sigma^{0}}{\sqrt{2}} & \Sigma^{+} & p \\
\Sigma^{-} &\frac{\Lambda}{\sqrt{6}}-\frac{\Sigma^{0}}{\sqrt{2}} & n \\
\Xi^{-} & \Xi^{0} & \frac{-2\Lambda }{\sqrt{6}}
\end{array}
\right),  \label{II-1}
\end{equation}
where $\Lambda$ and $\Sigma^{0}$ denote the un-mixed states, and the mixed states $\Lambda'$ and $\Sigma'^{0}$ will be introduced later.

Since $J/\psi$, $\Psi (2S)$, $\chi _{c0}$, $\chi _{c1}$ and $\chi _{c2}$ are all  SU(3) meson singlet,  the final effective interaction Hamiltonian and amplitude relations governing their decays into octet baryon pairs are similar.
According to the $J/\psi,\Psi (2S)\to \mathcal{B}_{8}\bar{\mathcal{B}}_{8}$ decays in Refs. \cite{Mo:2021asa,Ferroli:2020mra}, the final effective
interaction Hamiltonian for $\chi _{cJ}\rightarrow  \mathcal{B}_{8}\bar{\mathcal{B}}_{8}$ can be represented as
\begin{eqnarray}
\mathcal{H}_{eff}(\chi _{cJ}\rightarrow  \mathcal{B}_{8}\bar{\mathcal{B}}_{8})&=&  g_{0}\bar{\mathcal{B}}_{j}^{i}\mathcal{B}_{i}^{j}  +  g_{m}\big([\bar{\mathcal{B}}\mathcal{B}]_{f}\big)_{3}^{3}
                                                                       + g'_{m}\big([\bar{\mathcal{B}}\mathcal{B}]_{d}\big)_{3}^{3}  +  g_{e}\big([\bar{\mathcal{B}}\mathcal{B}]_{f}\big)_{1}^{1} + g'_{e}\big([\bar{\mathcal{B}}\mathcal{B}]_{d}\big)_{1}^{1},  \label{II-2}
\end{eqnarray}
with
\begin{eqnarray}
\big([\bar{\mathcal{B}}\mathcal{B}]_{f}\big)_{j}^{i}&=&\bar{\mathcal{B}}_{k}^{i}\mathcal{B}_{j}^{k}-\bar{\mathcal{B}}_{j}^{k}\mathcal{B}_{k}^{i},\label{II-4}\\
\big([\bar{\mathcal{B}}\mathcal{B}]_{d}\big)_{j}^{i}&=&\bar{\mathcal{B}}_{k}^{i}\mathcal{B}_{j}^{k}+\bar{\mathcal{B}}_{j}^{k}\mathcal{B}_{k}^{i}-\frac{2}{3}\delta _{j}^{i}(\bar{\mathcal{B}}_{n}^{m}\mathcal{B}_{m}^{n}).\label{II-5}
\end{eqnarray}
where $\mathcal{B}_{j}^{i}$ denotes the matrix element of the baryon  octet $\mathcal{B}_8$ with row $i$  and column $j$, $g_{0,m,e}$ and $g'_{m,e}$ are  nonperturbative coupling coefficients,
$g_{0}$ is the nonperturbative coefficient under the SU(3) flavor symmetry,  two types of the SU(3) flavor breaking effects are considered, and they are mass breaking $g_{m}$/$g'_{m}$ and  electromagnetic breaking $g_{e}$/$g'_{e}$.   Further details can be found in Refs. \cite{Mo:2021asa,Ferroli:2020mra,BaldiniFerroli:2019abd}.
Since these coefficients $g_{0,m,e}$ and $g'_{m,e}$ vary across different $\chi _{c0}$, $\chi _{c1}$ and $\chi _{c2}$ decays, we redefine them as  $A_J=g_{0}$, $D_J=g'_{e}/3$, $F_J=-g_{e}$, $D'_J=-g_{m}'/3$, and $F'_J=g_{m}$.
Using Eq. (\ref{II-2}), the  amplitudes
of $\chi _{cJ}\rightarrow \mathcal{B}_8\bar{\mathcal{B}}_8$ are parameterized,
and the results are listed  in the first nine lines of Tab. \ref{Tab:AmpB8B8}.

\begin{table}[b]
\renewcommand\arraystretch{1.2}
\tabcolsep 0.4in
\caption{ Amplitudes  for the $\chi _{cJ}\rightarrow \mathcal{B}_8\bar{\mathcal{B}}_8$ decays. $A_J=g_{0}$, $D_J=g'_{e}/3$, $F_J=-g_{e}$, $D'_J=-g_{m}'/3$, and $F'_J=g_{m}$. }
\begin{center}
\begin{tabular}{ll}
\hline\hline
Decay modes                                            & Amplitudes  \\  \hline
$\chi _{cJ}\rightarrow p\bar{p}$                                             & $A_J+D_J+F_J-D'_J+F'_J$                       \\
$\chi _{cJ}\rightarrow n\bar{n}$                                             & $A_J-2D_J-D'_J+F'_J$                        \\
$\chi _{cJ}\rightarrow\Sigma ^{+}\bar{\Sigma}^{-}$                          & $A_J+D_J+F_J+2D'_J$                                     \\
$\chi _{cJ}\rightarrow\Sigma ^{-}\bar{\Sigma}^{+}$                          & $A_J+D_J-F_J+2D'_J$                                     \\
$\chi _{cJ}\rightarrow\Xi ^{0}\bar{\Xi}^{0}$                                & $A_J-2D_J-D'_J-F'_J$                        \\
$\chi _{cJ}\rightarrow\Xi ^{-}\bar{\Xi}^{+}$                                & $A_J+D_J-F_J-D'_J-F'_J$                       \\
$\chi _{cJ}\rightarrow\Sigma^{0}\bar{\Sigma}^{0}$                           & $A_J+D_J+2D'_J$                                       \\
$\chi _{cJ}\rightarrow\Lambda\bar{\Lambda}$                                 & $A_J-D_J-2D'_J$                                       \\
$\chi _{cJ}\rightarrow\Sigma^{0}\bar{\Lambda}+c.c$                          & $\sqrt{3}D_J$                                                  \\  \hline
$\chi _{cJ}\rightarrow\Sigma'^{0}\bar{\Sigma}'^{0}$                         &   $A_J+(\cos 2\alpha -\sqrt{3}/2\sin 2\alpha )D_J+(2\cos 2\alpha )D'_J$ \\
$\chi _{cJ}\rightarrow\Lambda'\bar{\Lambda}'$                               &   $A_J+(\sqrt{3}/2\sin 2\alpha -\cos 2\alpha )D_J-(2\cos 2\alpha)D'_J$ \\
$\chi _{cJ}\rightarrow\Sigma'^{0}\bar{\Lambda}'+c.c $                       &   $(\sqrt{3}\cos 2\alpha+2\sin 2\alpha )D_J+(4\sin 2\alpha )D'_J$ \\ \hline

\end{tabular}
\end{center}\label{Tab:AmpB8B8}
\end{table}

In addition, in the case of $\Lambda$ and $\Sigma^{0}$ baryons, a small mixing angle $\alpha$ exists between the isoscalar state ($\Lambda$) and the neutral component of the isotriplet ($\Sigma^{0}$) owing to isospin violation. The mixing between $\Lambda$ and $\Sigma^{0}$ baryons have been studied in many works, for examples, in Refs. \cite{Dalitz:1964es,Karl:1994ie,Gal:2015iha,Geng:2020tlx}.
Then the  physical mass eigenstates $\Lambda'$ and $\Sigma'^{0}$ are
\begin{eqnarray}
\Lambda' &=&\Lambda\cos \alpha -\Sigma^{0}\sin \alpha,  \nonumber \\
\Sigma'^{0} &=&\Lambda\sin \alpha +\Sigma^{0}\cos \alpha. \label{II-6}
\end{eqnarray}
The  mixing angle $\alpha$ is estimated by different approaches,  for instance,
quark model \cite{Dalitz:1964es,Gal:1967oxu,Isgur:1979ed}, chiral perturbation theory \cite{Gasser:1982ap}, QCD sum rules in vacuum \cite{Yagisawa:2001gz,Zhu:1998ai}, QCD
sum rules \cite{Radici:2001pq,Aliev:2015ela}, and in lattice QCD \cite{Horsley:2014koa}. The mixing angle is predicted by these approaches as in the range of  $[0.14,2.0]\times10^{-2}$ radian.  After consider the $\Sigma^{0}-\Lambda$ mixing,  the  parameterized amplitudes
of $\chi _{cJ}\rightarrow \Sigma'^{0}\bar{\Sigma}'^{0},\Lambda' \bar{\Lambda}'$ and $\Sigma'^{0}\bar{\Lambda}'+\bar{\Sigma}'^{0}\Lambda'$ decays are given in the last three lines of Tab. \ref{Tab:AmpB8B8}.

It is important to note that there exist three decay modes: purely strong,
purely electromagnetic, and mixed strong-electromagnetic interactions in the studies of  $J/\psi $  and $\Psi (2S)$ decays \cite{BaldiniFerroli:2019abd,Ferroli:2020mra}, the amplitude can be broken down into
three parts: $A_{\mathcal{B}\bar{\mathcal{B}}}^{ggg}$, $A_{\mathcal{B}\bar{\mathcal{B}}}^{\gamma }$, and $A_{\mathcal{B}\bar{\mathcal{B}}}^{gg\gamma }$, corresponding to the three decay modes. The total amplitude $A_{\mathcal{B}\bar{\mathcal{B}}}=$
$A_{\mathcal{B}\bar{\mathcal{B}}}^{ggg}+A_{\mathcal{B}\bar{\mathcal{B}}}^{\gamma }+A_{\mathcal{B}\bar{\mathcal{B}}}^{gg\gamma }$ consists of the purely strong part $A_{\mathcal{B}\bar{\mathcal{B}}}^{ggg}=$ $\left( A,D',F'\right) e^{i\varphi }$, the purely
electromagnetic part $A_{\mathcal{B}\bar{\mathcal{B}}}^{\gamma }=\left( D,F\right) $, and the mixed strong-electromagnetic part $A_{\mathcal{B}\bar{\mathcal{B}}}^{gg\gamma }=A_{\mathcal{B}\bar{\mathcal{B}}}^{ggg}R$, where the phase $\varphi $ represents the relative phase between
the strong and electromagnetic amplitudes and the ratio $R$ denotes the relationship between $A_{\mathcal{B}\bar{\mathcal{B}}}^{gg\gamma }$ and $A_{\mathcal{B}\bar{\mathcal{B}}}^{ggg}$.
However, for the $\chi _{cJ}\rightarrow \mathcal{B}_8\bar{\mathcal{B}}_8$ decays, present experimental data are not enough to determine all seven coupling coefficients. Therefore, we focus on the dominant contributions of the purely strong and purely electromagnetic interactions, and neglect the weakest contribution part $A_{\mathcal{B}\bar{\mathcal{B}}}^{gg\gamma }$.
This leads us to consider the amplitude as $A(\chi _{cJ}\to \mathcal{B}_8\bar{\mathcal{B}}_8)=\left(A_J,D'_J,F'_J\right) e^{i\varphi_J }+\left( D_J,F_J\right) $,
which  only includes six coupling coefficients in each $\chi _{c0}$, $\chi _{c1}$ or $\chi _{c2}$ decay.

In terms of the decay amplitude of $\chi_{cJ}\rightarrow \mathcal{B}_8\bar{\mathcal{B}}_8$ given in Tab. \ref{Tab:AmpB8B8},    the branching ratio ($Br$) can be written as
\begin{equation}
Br(\chi _{cJ}\to \mathcal{B}_8\bar{\mathcal{B}}_8)=\frac{\left\vert \vec{p}\right\vert }{8\pi M_{\chi
_{cJ}}^{2}\Gamma _{\chi _{cJ}}}\left\vert A(\chi _{cJ}\to \mathcal{B}_8\bar{\mathcal{B}}_8)\right\vert ^{2},  \label{II-9}
\end{equation}
where $\Gamma _{\chi_{cJ}}$ is the width of the $\chi_{cJ}$\ meson, and  $|\vec{p}|\equiv\frac{\sqrt{[M_{\chi _{cJ}}^{2}-(m_{\mathcal{B}_8}^{2}+m_{\bar{\mathcal{B}}_8}^{2})][M_{\chi_{cJ}}^{2}-(m_{\mathcal{B}_8}^{2}-m_{\bar{\mathcal{B}}_8}^{2})]}}{2M_{\chi _{cJ}}}$.

\subsection{Numerical results and discussion}
Numerical results of  the $\chi _{cJ}\rightarrow \mathcal{B}_8\bar{\mathcal{B}}_8$ decays will be given in this subsection. The theoretical input parameters, such as the lifetimes and the masses, and the
experimental data within the $2\sigma(1\sigma)$ error(s) from PDG \cite{ParticleDataGroup:2022pth}
will be used to obtain our numerical analysis of the $\chi_{cJ}\rightarrow \mathcal{B}_8\bar{\mathcal{B}}_8$ decays.

Up to now, seven branching ratios in the $\chi _{c0}\rightarrow \mathcal{B}_8\bar{\mathcal{B}}_8$ decays, six branching ratios and an upper limit in the $\chi _{c1}\rightarrow \mathcal{B}_8\bar{\mathcal{B}}_8$ decays, and six branching ratios and an upper limit in the $\chi _{c2}\rightarrow \mathcal{B}_8\bar{\mathcal{B}}_8$ decays have been given by the experiments, which with $2\sigma$ errors are listed in the second columns of Tab. \ref{Tab:Brchi02S}, Tab. \ref{Tab:Brchi12S}, and Tab. \ref{Tab:Brchi22S}, respectively.
Using the experimental constraints and the theoretical expressions of the branching ratios, we perform a global fitting to determine allowed ranges of the coupling parameter values.
And then, relevant not-yet measured  branching ratios are predicted by the constrained coupling coefficients.  After using all relevant experimental constraints and the theoretical input parameters within the $1\sigma$ error,  some $\chi_{cJ}$ decays  fail to obtain results,  so we will analyse the results within the $1\sigma $ error and $2\sigma $ errors together.

Three cases are discussed for our numerical results. {\bf In $C_1$ case},  only the SU(3) flavor symmetry contribution are considered, the SU(3) flavor breaking  contribution and the $\Sigma^{0}-\Lambda$ mixing  are not considered,  $i.e.$, the coupling  coefficients $A_J$ are free parameters, and $D_J=D'_J=F_J=F'_J=0$.  {\bf In $C_2$ case},  both the SU(3) flavor symmetry and the SU(3) flavor breaking contributions are considered, and the $\Sigma^{0}-\Lambda$ mixing is not considered. We set $A_J>=0$ and other five parameters are free parameters (same in $C_{3}$ cases).
{\bf In $C_3$ case},  all the SU(3) flavor symmetry,  the SU(3) flavor breaking contributions, and the $\Sigma^{0}-\Lambda$ mixing are considered, and the mixing angle $\alpha= (0.015\pm0.001)$ radian are taken from the theoretical calculation by using PDG input masses \cite{Gal:2015iha}.

\begin{itemize}

\item{\bf Case $C_1$}: In this case, only one coupling parameter $A_J$ for the $\chi _{cJ}\rightarrow \mathcal{B}_8\bar{\mathcal{B}}_8$ decays. It is apparent that the range of $|A_J|$ can be directly inferred from experimental values and Eq. (\ref{II-9}). However, the permissible values of $A_J$  for each decay channel may differ significantly. We take $\chi _{c0}\rightarrow \mathcal{B}_8\bar{\mathcal{B}}_8$ as examples.
The central values of $|A_0|$  are $0.022,0.030,0.035,0.035,0.036,0.030,0.037$ from the experimental data
of the branching ratios of $\chi _{c0}\rightarrow p\bar{p},\Lambda \bar{\Lambda}, \Sigma ^{0}\bar{\Sigma}^{0},\Sigma ^{+}\bar{\Sigma}^{-}, \Sigma ^{-}\bar{\Sigma}^{+}, \Xi ^{0}\bar{\Xi}^{0}, \Xi ^{-}\bar{\Xi}^{+}$, respectively.
Seven central values are in the same order of magnitude, but some values are obviously different.  We can not find one value of $|A_0|$,  which  satisfy all present relevant experiential data within $2\sigma$ error bars.
Therefore, we only use $Br(\chi _{cJ}\rightarrow p\bar{p})$ to constrain $|A_J|$, and then use the constrained $|A_J|$ to give the branching ratios of other processes.
The results within the $2\sigma$ errors ($1\sigma$ error)  are listed in the third columns of  Tab. \ref{Tab:Brchi02S}, Tab. \ref{Tab:Brchi12S}, and Tab. \ref{Tab:Brchi22S} (Tab. \ref{Tab:Brchi01S}, Tab. \ref{Tab:Brchi11S}, and Tab. \ref{Tab:Brchi21S}).
It becomes apparent that these SU(3) flavor symmetry predictions do not effectively align with the experimental values, and the  SU(3) flavor breaking effects must be considered. These SU(3) flavor symmetry predictions can serve as benchmarks for gauging the magnitude of the breaking effect.
Additionally, there are still two decay branching ratios  which no experimental data is available,
and they are  predicted by the constrained $|A_J|$ from the data of $Br(\chi _{cJ}\rightarrow p\bar{p})$.

\item{\bf Case $C_2$}: We conduct a global fit employing the six free
parameters $\left( A_{J},D_{J},F_{J},D_{J}',F_{J}',\varphi _{J}\right) $ of the SU(3) amplitude without the mixing angle $\alpha$.
The results with $2\sigma $ errors are provided in the fourth columns of Tab. \ref{Tab:Brchi02S}, Tab. \ref{Tab:Brchi12S}, and Tab. \ref{Tab:Brchi22S}, which satisfy all of the existing relevant experimental data.
Comparing our predictions with corresponding experimental data, one can see many of them are same, which  means that both upper limits and lower limits of these data give effective constraints on relevant parameters.
If only upper limit or lower limit of the prediction is same as its experimental one, that means only upper limit or lower limit gives effective constraint. One can similarly analyze in other cases.
For the $\chi _{c0}\rightarrow \mathcal{B}_8\bar{\mathcal{B}}_8$ decays given in the fourth column of  Tab. \ref{Tab:Brchi02S},  one can see
that both the  experimental upper and lower limits  of $Br(\chi_{c0}\to p\bar{p})$,
$Br(\chi_{c0}\to \Lambda \bar{\Lambda})$,
$Br(\chi_{c0}\to \Sigma^{0}\bar{\Sigma}^{0})$,
$Br(\chi_{c0}\to \Sigma ^{-}\bar{\Sigma}^{+})$ and
$Br(\chi_{c0}\to \Xi ^{-}\bar{\Xi}^{+})$,  the  experimental  lower limit of
$Br(\chi_{c0}\to \Sigma ^{+}\bar{\Sigma}^{-})$  as well as the  experimental  upper limit of  $Br(\chi_{c0}\to \Xi ^{0}\bar{\Xi}^{0})$
give the effective constraints on the parameters $\left( A_{0},D_{0},F_{0},D_{0}',F_{0}'\right)$.
$Br(\chi_{c0}\to n\bar{n})$ are predicted in the order of $10^{-4}$. Due to weak constraint on $|D_0|$, the predicted $Br(\chi_{c0}\to \Lambda \bar{\Sigma}^{0}+c.c.)$ has the form $a\pm a$, the measurement of  $Br(\chi_{c0}\to \Lambda \bar{\Sigma}^{0}+c.c.)$ in the future  may give better constraint on $|D_0|$.
For the $\chi _{c1}\rightarrow \mathcal{B}_8\bar{\mathcal{B}}_8$ decays given in the fourth column of  Tab. \ref{Tab:Brchi12S},  one can see
that all measured branching ratios give the effective constraints on $\left( A_{1},D_{1},F_{1},D_{1}',F_{1}'\right)$. Both $Br(\chi_{c1}\to n\bar{n})$ and $Br(\chi_{c1}\to \Lambda \bar{\Sigma}^{0}+c.c.)$ are predicted in the form of $a\pm a$, and the prediction of $Br(\chi_{c1}\to n\bar{n})$ may be quite larger than its SU(3) flavor symmetry prediction. Because of the presence of another numerical range for $D_{1}'$ and $F_{1}'$, $Br(\chi_{c1}\to n\bar{n})$ displays another quite larger branching ratio.
As for the $\chi _{c2}\rightarrow \mathcal{B}_8\bar{\mathcal{B}}_8$ decays given in the fourth column of  Tab. \ref{Tab:Brchi22S},  all measured branching ratios except the lower limit of $Br(\chi_{c2}\to \Xi ^{0}\bar{\Xi}^{0})$
give effective constraints on  $\left( A_{2},D_{2},F_{2},D_{2}',F_{2}'\right)$. It can be observed that the phase angles $\varphi _{J}$ are not evidently constrained by current experimental data, primarily due to the limited availability of experimental data and the lack of additional parameter constraints. Moreover, some errors of the constrained $A_{J},D_{J},F_{J},D_{J}',F_{J}'$ are large.

The results with $1\sigma $ errors are listed in the fourth  columns of Tab. \ref{Tab:Brchi01S}, Tab. \ref{Tab:Brchi11S}, and Tab. \ref{Tab:Brchi21S}.  For the $\chi _{c0}\rightarrow \mathcal{B}_8\bar{\mathcal{B}}_8$ decays, the upper limit of $Br(\chi_{c0}\to p\bar{p})$ is not used to constrain the relevant parameters and give the branching ratio predictions. One can see that predicted $Br(\chi_{c0}\to p\bar{p})$ is much larger than its experimental upper limit. And all errors of the parameter $\left( A_{J},D_{J},F_{J},D_{J}',F_{J}'\right)$ are still very large.

\item{\bf Case $C_3$}: If considering  the $\Sigma^0-\Lambda$ mixing angle and setting the mixing angle $\alpha=0.015\pm0.001$ radian \cite{Gal:2015iha},  all allowed coupling parameter spaces and the branching ratios are recalculated, and the results  with $2\sigma$ ($1\sigma$) errors are listed  in the fifth columns of   Tab. \ref{Tab:Brchi02S}, Tab. \ref{Tab:Brchi12S}, and Tab. \ref{Tab:Brchi22S} ( Tab. \ref{Tab:Brchi01S}, Tab. \ref{Tab:Brchi11S}, and Tab. \ref{Tab:Brchi21S}).  Since the mixing angle $\alpha$ is quite small, the allowed parameter spaces and the branching ratio predictions are similar to ones in the $C_2$ case.

\end{itemize}

From the results in   Tabs. \ref{Tab:Brchi02S}-\ref{Tab:Brchi21S}, one can see some absolute values of the constrained  $D_{J},F_{J},D_{J}'$ or $F_{J}'$ are not far less than   that for $A_{J}$, which means the SU(3) flavor breaking effects (both charge braking $D_J,F_J$ and mass breaking $D'_J,F'_J$) are not constrained much for $\chi _{cJ}\rightarrow \mathcal{B}_8\bar{\mathcal{B}}_8$ decays by present experimental measurements.
Referring to charmonium $J/\psi,\Psi(2S)\rightarrow \mathcal{B}_8\bar{\mathcal{B}}_8$ decays in Refs. \cite{Ferroli:2020mra,Mo:2023wrf}, the maximum value of   the constrained  $|D/A|,|F/A|,|D'/A|$ and $|F'/A|$ is 17.92\%.
Next, we  assume that the ratio $|D_J/A_J|,|F_J/A_J|,|D'_J/A_J|$ and $|F'_J/A_J|$ are less than or equal to 20\% to obtain our numerical results, then the $C_{2,3}$ cases are renamed $C'_{2,3}$ cases.

The results of the  $\chi _{c0}\rightarrow \mathcal{B}_8\bar{\mathcal{B}}_8$ decays within the $2\sigma$ errors ($1\sigma$ error) are listed in the last two columns of  Tab. \ref{Tab:Brchi02S} (Tab. \ref{Tab:Brchi01S}).
One can see that the predicted $Br(\chi_{c0}\to p\bar{p})$ in the $C'_{2,3}$ cases are similar to ones in the $C'_{2,3}$ cases, nevertheless, the  predicted $Br(\chi_{c0}\to \Lambda^{(')} \bar{\Sigma}^{(')0}+c.c.)$ in the $C'_{2,3}$ cases have smaller allowed spaces.

As for the $\chi _{c1,2}\rightarrow \mathcal{B}_8\bar{\mathcal{B}}_8$ decays, the results are failed to obtain after assuming all $|D_J/A_J|,|F_J/A_J|,|D'_J/A_J|$ and $|F'_J/A_J|$ are less than or equal to 20\%.
The difficulty lies in $|D'_J/A_J|\leq20\%$,  if we remove the limit of $|D'_J/A_J|\leq20\%$, the results within the $2\sigma$ errors can be obtained, nevertheless, $|D'_{1,2}|$ are really large.
Then the results of the  $\chi _{c1,2}\rightarrow \mathcal{B}_8\bar{\mathcal{B}}_8$ decays within the $2\sigma$ errors are listed in the last two columns of  Tab. \ref{Tab:Brchi12S} and Tab. \ref{Tab:Brchi22S}. One can find that the branching ratios predictions in the $C'_{2,3}$ cases, specially for  $Br(\chi_{c0}\to n\bar{n})$  and $Br(\chi_{c0}\to \Lambda \bar{\Sigma}^{0}+c.c.)$,  are more accurate than ones in the $C_{2,3}$ cases.
If using the experimental constraints and theoretical inputs within $1\sigma$ error, it is failed to obtain the results, so we ignore the lower limits of $Br(\chi_{c1}\to \Sigma^{(')0}\bar{\Sigma}^{(')0})$,  $Br(\chi_{c1}\to \Xi ^{-}\bar{\Xi}^{+})$ and $Br(\chi_{c2}\to \Xi ^{-}\bar{\Xi}^{+})$ to get the results within $1\sigma$ error, which are  given in the last two columns of  Tab. \ref{Tab:Brchi11S} and Tab. \ref{Tab:Brchi21S}.

\begin{table}[htb]
\caption{The allowed amplitude parameters and branching ratios (in units of $10^{-4}$) of $\protect\chi_{c0}\rightarrow \mathcal{B}_8\bar{\mathcal{B}}_8$ decays  within $2\sigma $ errors.
{\bf $C_1$ case}: only the SU(3) flavor symmetry contribution are considered. {\bf $C_2$ case}:  both the SU(3) flavor symmetry and the SU(3) flavor breaking contributions are considered, and the $\Sigma^{0}-\Lambda$ mixing is not considered. {\bf $C_3$ case}:  all the SU(3) flavor symmetry,  the SU(3) flavor breaking contributions, and the $\Sigma^{0}-\Lambda$ mixing are considered, and $\alpha = (0.015\pm0.001)$ radians \cite{Gal:2015iha}. $^\dag$Only its data are used to determine the relevant parameter $|A_J|$. }
\centering%
\renewcommand\arraystretch{0.8}\tabcolsep 0.03in
{\scriptsize
\begin{tabular}{lcccc|cc}\hline\hline
                                                                                             & Exp. data    & \begin{tabular}{c} Predictions\\  in $C_1$ case \end{tabular}&\begin{tabular}{c}Predictions  \\in  $C_2$ case \end{tabular}&\begin{tabular}{c}Predictions   \\ in $C_3$ case  \end{tabular}&\begin{tabular}{c}Predictions   \\ in $C'_2$ case  \end{tabular} &\begin{tabular}{c}Predictions   \\ in $C'_3$ case  \end{tabular}\\ \hline
$Br(\chi_{c0}\to p\bar{p})$                                                                  & $2.21\pm 0.16$           & $2.21\pm 0.16^\dag$  & $2.21\pm 0.16$            & $2.21\pm 0.16$                                  & $2.21\pm 0.16$           & $2.21\pm 0.16$                           \\
$Br(\chi_{c0}\to n\bar{n})$                                                                  & $\cdots$                 & $2.21\pm 0.16$       & $3.77\pm 2.40$            & $3.77\pm 2.37$                                  & $3.98\pm 2.00$           & $3.97\pm 1.99$                             \\
$Br(\chi_{c0}\to \Lambda^{(')} \bar{\Lambda}^{(')})$                                         & $3.59\pm 0.30$           & $2.00\pm 0.14$       & $3.59\pm 0.30$            & $3.59\pm 0.30$                                  & $3.59\pm 0.30$           & $3.59\pm 0.30$                             \\
$Br(\chi_{c0}\to \Sigma^{(')0}\bar{\Sigma}^{(')0})$                                          & $4.68\pm 0.64$           & $1.89\pm 0.14$       & $4.68\pm 0.64$            & $4.68\pm 0.64$                                  & $4.67\pm 0.63$           & $4.66\pm 0.62$                             \\
$Br(\chi_{c0}\to \Sigma ^{+}\bar{\Sigma}^{-})$                                               & $4.6\pm 1.6$             & $1.89\pm 0.14$       & $3.92\pm 0.92$            & $3.92\pm 0.92$                                  & $3.91\pm 0.91$           & $3.94\pm 0.94$                            \\
$Br(\chi_{c0}\to \Sigma ^{-}\bar{\Sigma}^{+})$                                               & $5.1\pm 1.0$             & $1.89\pm 0.14$       & $5.10\pm 1.00$            & $5.10\pm 1.00$                                 & $5.10\pm 1.00$           & $5.10\pm 1.00$                            \\
$Br(\chi_{c0}\to \Xi ^{0}\bar{\Xi}^{0})$                                                     & $3.1\pm 1.6$             & $1.69\pm 0.12$       & $4.03\pm 0.67$            & $4.03\pm 0.67$                                  & $4.04\pm 0.66$           & $4.05\pm 0.65$                            \\
$Br(\chi_{c0}\to \Xi ^{-}\bar{\Xi}^{+})$                                                     & $4.8\pm 1.4$             & $1.67\pm 0.12$       & $4.80\pm 1.40$            & $4.80\pm 1.40$                                  & $4.80\pm 1.40$           & $4.80\pm 1.40$                            \\
$Br(\chi_{c0}\to \Lambda^{(')} \bar{\Sigma}^{(')0}+c.c.)$                                    & $\cdots$                 & $\cdots$             & $0.37\pm 0.37$            & $0.80\pm 0.80$                                & $0.13\pm 0.13$           & $0.26\pm 0.26$                            \\ \hline
$A_{0}$ $(10^{-2}$ GeV$)$                                                                    & $\cdots$                 & $2.21\pm 0.20$       & $3.14\pm0.33 $            & $3.13\pm 0.34$                                  & $3.17\pm0.29 $           & $3.17\pm 0.29$                            \\
$D_{0}$ $(10^{-3}$ GeV$)$                                                                    & $\cdots$                 & $\cdots$             & $0.20\pm 7.70$            & $-0.01\pm7.91$                                 & $-0.10\pm 4.74$          & $-0.10\pm4.74$                             \\
$F_{0}$ $(10^{-3}$ GeV$)$                                                                    & $\cdots$                 & $\cdots$             & $0.32\pm 13.92$           & $-0.16\pm13.44$                                 & $-0.09\pm 6.69$          & $-0.09\pm6.69$                             \\
$D_{0}'$ $(10^{-3}$ GeV$)$                                                                   &$\cdots$                  &$\cdots$              & $1.67\pm 1.70$            & $1.64\pm1.74$                                   & $1.74\pm 1.74$           & $1.73\pm1.75$                           \\
$F_{0}'$ $(10^{-3}$ GeV$)$                                                                   &$\cdots$                  & $\cdots$             & $-4.44\pm5.28$            & $-4.61\pm5.45$                                 & $-2.95\pm3.89$           & $-2.95\pm3.89$                            \\
$\varphi _{0}$    (radian)                                                                   &$\cdots$                  &$\cdots$              & $0\pm\pi$                 & $0\pm\pi$                                            & $0\pm\pi$                & $0\pm\pi$                                 \\
$\alpha$       (radian)                                                                      &$\cdots$                  &$\cdots$              & $0$                       & $0.015\pm0.002$                                  & $0$                      & $0.015\pm0.002$                          \\ \hline
\end{tabular}}\label{Tab:Brchi02S}
%
\caption{The allowed amplitude parameters and branching ratios (in units of $10^{-5}$) of $\protect\chi_{c1}\rightarrow \mathcal{B}_8\bar{\mathcal{B}}_8$ decays within $2\sigma $ errors. }
\centering%
\renewcommand\arraystretch{0.8}\tabcolsep 0.03in
{\scriptsize
\begin{tabular}{lcccc|cc}\hline\hline
              & Exp. data  & \begin{tabular}{c} Predictions\\ in $C_1$ case \end{tabular}&\begin{tabular}{c}Predictions   \\ in $C_2$ case \end{tabular}&\begin{tabular}{c}Predictions \\in  $C_3$ case  \end{tabular} &\begin{tabular}{c}Predictions \\in  $C'_2$ case  \end{tabular}  &\begin{tabular}{c}Predictions \\in  $C'_3$ case  \end{tabular}               \\ \hline
$Br(\chi_{c1}\to p\bar{p})$                                                                       &   $7.60\pm 0.68$             & $7.60\pm 0.68^\dag$     &    $7.60\pm 0.68$                 &    $7.60\pm 0.68$                                                  &    $7.60\pm 0.68$                   &    $7.60\pm 0.68$                             \\
$Br(\chi_{c1}\to n\bar{n})$                                                                       &   $\cdots$                   & $7.60\pm 0.68$       & \begin{tabular}{c}  $7.40\pm7.40$ \\or $56.06\pm24.14$\end{tabular}         &  \begin{tabular}{c}$7.57\pm 7.57$ \\ or $52.66\pm23.67$\end{tabular}                 &    $5.75\pm1.90$                    &    $5.79\pm 1.81$                               \\
$Br(\chi_{c1}\to \Lambda^{(')} \bar{\Lambda}^{(')})$                                              &   $12.7\pm 1.6$              & $6.94\pm 0.62$          &    $12.70\pm 1.60$                &    $12.70\pm 1.60$                                                 &    $12.99\pm 1.31$                  &    $12.99\pm 1.31$                               \\
$Br(\chi_{c1}\to \Sigma^{(')0}\bar{\Sigma}^{(')0})$                                               &   $4.2\pm 1.2$               & $6.60\pm 0.59$          &    $4.20\pm 1.20$                 &    $4.20\pm 1.20$                                                   &    $4.20\pm 1.20$                   &    $4.18\pm 1.18$                                 \\
$Br(\chi_{c1}\to \Sigma ^{+}\bar{\Sigma}^{-})$                                                    &   $3.6\pm 1.4$               & $6.59\pm 0.59$          &    $3.60\pm 1.40$                 &    $3.60\pm 1.40$                                                   &    $3.71\pm 1.29$                   &    $3.71\pm 1.29$                                 \\
$Br(\chi_{c1}\to \Sigma ^{-}\bar{\Sigma}^{+})$                                                    &   $5.7\pm 3.0$               & $6.59\pm 0.59$          &    $5.70\pm 3.00$                 &    $5.70\pm 3.00$                                                   &    $4.33\pm 1.63$                   &    $4.33\pm 1.63$                                \\
$Br(\chi_{c1}\to \Xi ^{0}\bar{\Xi}^{0})$                                                          &   $<6$                       & $5.96\pm 0.53$          &    $3.52\pm 2.48$                 &    $3.52\pm 2.48$                                                  &    $3.77\pm 1.21$                   &    $3.76\pm 1.12$                                \\
$Br(\chi_{c1}\to \Xi ^{-}\bar{\Xi}^{+})$                                                          &   $8.0\pm 4.2$               & $5.92\pm 0.53$          &    $8.00\pm 4.20$                 &    $8.00\pm 4.20$                                                   &    $5.19\pm 1.39$                   &    $5.23\pm 1.43$                                 \\
$Br(\chi_{c1}\to \Lambda^{(')} \bar{\Sigma}^{(')0}+c.c.)$                                         &   $\cdots$                   & $\cdots$                &    $6.34\pm 6.34$                 &    $11.84\pm 11.83$                                             &    $0.06\pm 0.06$                   &    $0.18\pm0.18$                                \\ \hline
$A_{1}$ $(10^{-3}$ GeV$)$                                                                         &   $\cdots$                   & $3.65\pm 0.33$          &    $1.09\pm0.43$                  &    $1.09\pm 0.43$                                                  &    $1.17\pm0.28$                    &    $1.17\pm 0.28$                                \\
$D_{1}$ $(10^{-3}$ GeV$)$                                                                         &   $\cdots$                   & $\cdots$                &    $0.01\pm 2.92$                 &    $0.01\pm 2.92$                                                  &    $0.00\pm 0.28$                   &    $0.00\pm 0.28$                             \\
$F_{1}$ $(10^{-3}$ GeV$)$                                                                         &   $\cdots$                   & $\cdots$                &    $0.03\pm 2.67$                 &    $0.03\pm 2.67$                                                   &    $0.00\pm 0.28$                   &    $0.00\pm 0.28$                                \\
$D_{1}'$ $(10^{-3}$ GeV$)$                                                                        &   $\cdots$                   & $\cdots$                & \begin{tabular}{c} $-2.09\pm 0.58$\\or $-0.82\pm0.41$ \end{tabular}                   & \begin{tabular}{c} $-2.08\pm0.57$\\or $-0.80\pm0.42$ \end{tabular}                  &    $-2.06\pm 0.25$                  &    $-2.07\pm0.25$                               \\
$F_{1}'$ $(10^{-3}$ GeV$)$                                                                        &   $\cdots$                   & $\cdots$                &  \begin{tabular}{c}$-0.42\pm1.67$\\ or $3.47\pm1.25$ \end{tabular}                     &  \begin{tabular}{c}$-0.42\pm1.67$\\ or $3.45\pm1.26$ \end{tabular}                         &    $0.06\pm0.22$                    &    $0.06\pm0.22$                                 \\
$\varphi _{1}$ (radian)                                                                           &   $\cdots$                   & $\cdots$                &    $0\pm\pi$                      &    $0\pm\pi$                   &    $0\pm\pi$                        &    $0\pm\pi$                                        \\
$\alpha$   (radian)                                                                               &   $\cdots$                   & $\cdots$                &    $0$                            &    $0.015\pm0.002$                                                  &    $0$                              &    $0.015\pm0.002$                               \\ \hline
\end{tabular}}\label{Tab:Brchi12S}
%
\caption{The allowed amplitude parameters and branching ratios (in units of $10^{-5}$) of $\protect\chi_{c2}\rightarrow \mathcal{B}_8\bar{\mathcal{B}}_8$ decays within $2\sigma $ errors. }
\centering%
\renewcommand\arraystretch{0.8}\tabcolsep  0.03in
{\scriptsize
\begin{tabular}{lcccc|cc}\hline\hline
              & Exp. data & \begin{tabular}{c} Predictions \\ in $C_1$ case\end{tabular}&\begin{tabular}{c}Predictions  \\ in  $C_2$ case\end{tabular}&\begin{tabular}{c}Predictions  \\ in  $C_3$ case \end{tabular}&\begin{tabular}{c}Predictions  \\ in  $C'_2$ case \end{tabular} &\begin{tabular}{c}Predictions  \\ in  $C'_3$ case \end{tabular}\\ \hline
$Br(\chi_{c2}\to p\bar{p})$                                         &    $7.33\pm 0.66$         &   $7.33\pm 0.66^\dag$       &     $7.33\pm 0.66$             &      $7.33\pm 0.66$                                            &     $7.33\pm 0.66$        &      $7.33\pm 0.66$                                \\
$Br(\chi_{c2}\to n\bar{n})$                                         &    $\cdots$               &   $7.33\pm 0.66$            &     $7.31\pm 7.31$             &      $7.31\pm 7.31$                                           &     $8.26\pm 4.04$        &      $8.23\pm 3.79$                                \\
$Br(\chi_{c2}\to \Lambda^{(')} \bar{\Lambda}^{(')})$                &    $18.3\pm 3.2$          &   $6.72\pm 0.60$            &     $18.30\pm 3.20$            &     $18.30\pm 3.20$                                          &     $18.32\pm 3.18$       &      $18.35\pm 3.15$                              \\
$Br(\chi_{c2}\to \Sigma^{(')0}\bar{\Sigma}^{(')0})$                 &    $3.7\pm 1.2$           &   $6.40\pm 0.58$            &     $3.70\pm 1.20$             &      $3.70\pm 1.20$                                            &     $3.70\pm 1.20$        &      $3.69\pm 1.19$                               \\
$Br(\chi_{c2}\to \Sigma ^{+}\bar{\Sigma}^{-})$                      &    $3.4\pm 1.4$           &   $6.40\pm 0.58$            &     $3.40\pm 1.40$             &      $3.40\pm 1.40$                                             &     $3.51\pm 1.29$        &      $3.51\pm 1.29$                                \\
$Br(\chi_{c2}\to \Sigma ^{-}\bar{\Sigma}^{+})$                      &    $4.4\pm 3.6$           &   $6.40\pm 0.58$            &     $4.52\pm 3.48$             &      $4.57\pm 3.43$                                            &     $3.38\pm 1.66$        &      $3.36\pm 1.67$                                 \\
$Br(\chi_{c2}\to \Xi ^{0}\bar{\Xi}^{0})$                            &    $<10$                  &   $5.81\pm 0.52$            &     $6.36\pm 3.64$             &      $6.52\pm 3.48$                                           &     $7.03\pm 2.97$        &      $7.11\pm 2.89$                                \\
$Br(\chi_{c2}\to \Xi ^{-}\bar{\Xi}^{+})$                            &    $14.2\pm 6.4$          &   $5.77\pm 0.52$            &     $14.20\pm 6.40$            &      $14.20\pm6.40$                                           &     $9.93\pm 2.13$        &      $9.93\pm2.13$                               \\
$Br(\chi_{c2}\to \Lambda^{(')} \bar{\Sigma}^{(')0}+c.c)$            &    $\cdots$               &   $\cdots$                  &     $3.65\pm 3.65$             &      $7.25\pm 7.25$                                            &     $0.12\pm 0.12$        &      $0.27\pm 0.27$                                \\ \hline
$A_{2}$ $(10^{-3}$  GeV$)$                                          &    $\cdots$               &   $5.52\pm 0.50$            &     $2.54\pm 0.84$             &      $2.54\pm 0.84$                                           &     $2.69\pm 0.64$        &      $2.69\pm 0.63$                                 \\
$D_{2}$ $(10^{-3}$  GeV$)$                                          &    $\cdots$               &   $\cdots$                  &     $-0.09\pm 3.39$             &      $-0.09\pm3.39$                                          &     $0.00\pm 0.61$        &      $0.00\pm 0.61$                                \\
$F_{2}$ $(10^{-3}$   GeV$)$                                         &    $\cdots$               &   $\cdots$                  &     $0.01\pm 4.15$             &      $0.01\pm 4.15$                                            &     $0.01\pm 0.65$        &      $0.01\pm 0.64$                                 \\
$D_{2}'$ $(10^{-3}$  GeV$)$                                         &   $\cdots$                &   $\cdots$                  &     $-3.49\pm 0.87$            &      $-3.51\pm0.89$                                            &     $-3.37\pm 0.41$       &      $-3.39\pm0.40$                                \\
$F_{2}'$ $(10^{-3}$   GeV$)$                                        &    $\cdots$               &   $\cdots$                  &     $-1.67\pm 2.44$            &      $-1.67\pm2.44$                                            &     $-0.21\pm 0.44$       &      $-0.21\pm0.43$                               \\
$\varphi _{2}$ (radian)                                             &    $\cdots$               &   $\cdots$                  &     $0\pm\pi$                  &      $0\pm\pi$                                                     &     $0\pm\pi$             &      $0\pm\pi$                                          \\
$\alpha$   (radian)                                                 &    $\cdots$               &   $\cdots$                  &     $0$                        &      $0.015\pm0.002$                                           &     $0$                   &      $0.015\pm0.002$                              \\ \hline
\end{tabular} }\label{Tab:Brchi22S}
\end{table}

\begin{table}[htb]
\caption{The allowed amplitude parameters and branching ratios (in units of $10^{-4}$) of $\protect\chi_{c0}\rightarrow \mathcal{B}_8\bar{\mathcal{B}}_8$ decays  within $1\sigma $ errors.
$^\dag$Only this data are used to determine the relevant parameter $|A_J|$. $^{U(L)}$Its experimental upper(lower) limit is not  used to determine the relevant parameters. The same in Tabs. \ref{Tab:Brchi11S}-\ref{Tab:Brchi21S}. }
\centering%
\renewcommand\arraystretch{0.8}\tabcolsep 0.03in
{\scriptsize
\begin{tabular}{lcccc|cc}\hline\hline
                                                                                             & Exp. data    & \begin{tabular}{c} Predictions\\  in $C_1$ case \end{tabular}&\begin{tabular}{c}Predictions  \\in  $C_2$ case \end{tabular}&\begin{tabular}{c}Predictions   \\ in $C_3$ case  \end{tabular} &\begin{tabular}{c}Predictions   \\ in $C'_2$ case  \end{tabular} &\begin{tabular}{c}Predictions   \\ in $C'_3$ case  \end{tabular} \\ \hline
$Br(\chi_{c0}\to p\bar{p})$                                                                  & $2.21\pm 0.08$           & $2.21\pm 0.08^\dag$  & $4.48\pm 1.81^U$          & $4.42\pm 1.75^U$                              & $4.41\pm 1.68^U$         & $4.39\pm 1.67^U$                           \\
$Br(\chi_{c0}\to n\bar{n})$                                                                  & $\cdots$                 & $2.21\pm 0.08$       & $3.86\pm 1.98$            & $3.96\pm 1.89$                                  & $3.31\pm 1.34$           & $3.32\pm 1.33$                           \\
$Br(\chi_{c0}\to \Lambda^{(')} \bar{\Lambda}^{(')})$                                         & $3.59\pm 0.15$           & $2.00\pm 0.07$       & $3.59\pm 0.15$            & $3.59\pm 0.15$                                  & $3.59\pm 0.15$           & $3.59\pm 0.15$                           \\
$Br(\chi_{c0}\to \Sigma^{(')0}\bar{\Sigma}^{(')0})$                                          & $4.68\pm 0.32$           & $1.89\pm 0.07$       & $4.68\pm 0.32$            & $4.68\pm 0.32$                                  & $4.68\pm 0.32$           & $4.68\pm 0.32$                           \\
$Br(\chi_{c0}\to \Sigma ^{+}\bar{\Sigma}^{-})$                                               & $4.6\pm 0.8$             & $1.89\pm 0.07$       & $4.60\pm 0.80$            & $4.60\pm 0.80$                                  & $4.60\pm 0.80$           & $4.57\pm 0.77$                             \\
$Br(\chi_{c0}\to \Sigma ^{-}\bar{\Sigma}^{+})$                                               & $5.1\pm 0.5$             & $1.89\pm 0.07$       & $5.10\pm 0.50$            & $5.10\pm 0.50$                                  & $5.10\pm 0.50$           & $5.10\pm 0.50$                             \\
$Br(\chi_{c0}\to \Xi ^{0}\bar{\Xi}^{0})$                                                     & $3.1\pm 0.8$             & $1.69\pm 0.06$       & $3.10\pm 0.80$            & $3.10\pm 0.80$                                  & $3.10\pm 0.80$           & $3.10\pm 0.80$                             \\
$Br(\chi_{c0}\to \Xi ^{-}\bar{\Xi}^{+})$                                                     & $4.8\pm 0.7$             & $1.67\pm 0.06$       & $4.80\pm 0.70$            & $4.80\pm 0.70$                                  & $4.80\pm 0.70$           & $4.80\pm 0.70$                           \\
$Br(\chi_{c0}\to \Lambda^{(')} \bar{\Sigma}^{(')0}+c.c.)$                                    & $\cdots$                 & $\cdots$             & $1.05\pm 1.05$            & $1.00\pm 1.00$                                 & $0.25\pm 0.25$           & $0.24\pm 0.24$                             \\ \hline
$A_{0}$ $(10^{-2}$ GeV$)$                                                                    & $\cdots$                 & $2.21\pm 0.10$       & $3.14\pm0.25 $            & $3.13\pm 0.26$                                & $3.20\pm0.19 $           & $3.20\pm 0.19$                             \\
$D_{0}$ $(10^{-3}$ GeV$)$                                                                    & $\cdots$                 & $\cdots$             & $-0.10\pm 13.36$          & $-0.11\pm13.17$                                & $0.04\pm 6.60$           & $0.05\pm6.58$                              \\
$F_{0}$ $(10^{-3}$ GeV$)$                                                                    & $\cdots$                 & $\cdots$             & $0.08\pm 16.34$           & $0.08\pm16.34$                                & $0.00\pm 6.63$           & $0.00\pm6.63$                              \\
$D_{0}'$ $(10^{-3}$ GeV$)$                                                                   &$\cdots$                  &$\cdots$              & $0.82\pm 0.92$            & $0.82\pm0.92$                                   &$0.83\pm 0.93$            & $0.83\pm0.93$                             \\
$F_{0}'$ $(10^{-3}$ GeV$)$                                                                   &$\cdots$                  & $\cdots$             & $-2.25\pm4.25$            & $-1.90\pm3.89$                                 & $-2.67\pm3.67$           & $-2.67\pm3.67$                            \\
$\varphi _{0}$    (radian)                                                                   &$\cdots$                  &$\cdots$              & $0\pm\pi$                 & $0\pm\pi$                                            & $0\pm\pi$                & $0\pm\pi$                                    \\
$\alpha$       (radian)                                                                      &$\cdots$                  &$\cdots$              & $0$                       & $0.015\pm0.001$                                 & $0$                      & $0.015\pm0.001$                           \\ \hline
\end{tabular}}\label{Tab:Brchi01S}
%
\caption{The allowed amplitude parameters and branching ratios (in units of $10^{-5}$) of $\protect\chi_{c1}\rightarrow \mathcal{B}_8\bar{\mathcal{B}}_8$ decays within $1\sigma $ errors. }
\centering%
\renewcommand\arraystretch{0.8}\tabcolsep 0.03in
{\scriptsize
\begin{tabular}{lcccc|cc}\hline\hline
              & Exp. data  & \begin{tabular}{c} Predictions\\ in $C_1$ case \end{tabular}&\begin{tabular}{c}Predictions   \\ in $C_2$ case \end{tabular}&\begin{tabular}{c}Predictions \\in  $C_3$ case  \end{tabular}&\begin{tabular}{c}Predictions \\in  $C'_2$ case  \end{tabular}  &\begin{tabular}{c}Predictions \\in  $C'_3$ case  \end{tabular}  \\ \hline
$Br(\chi_{c1}\to p\bar{p})$                                                                       &   $7.60\pm 0.34$             & $7.60\pm 0.34^\dag$     &    $7.60\pm 0.34$                 &    $7.60\pm 0.34$                                                  &    $7.56\pm 0.30$                   &    $7.60\pm 0.33$                               \\
$Br(\chi_{c1}\to n\bar{n})$                                                                       &   $\cdots$                   & $7.60\pm 0.34$          & \begin{tabular}{c}$6.35\pm5.68$\\ or $54.86\pm9.85$ \end{tabular}                     & \begin{tabular}{c} $6.14\pm5.88$\\  or $55.01\pm8.79$ \end{tabular}                             &    $5.20\pm0.55$                    &    $5.13\pm 0.69$                            \\
$Br(\chi_{c1}\to \Lambda^{(')} \bar{\Lambda}^{(')})$                                              &   $12.7\pm 0.8$              & $6.94\pm 0.31$          &    $12.70\pm 0.80$                &    $12.70\pm 0.80$                                                 &    $13.24\pm 0.26$                  &    $13.19\pm 0.31$                               \\
$Br(\chi_{c1}\to \Sigma^{(')0}\bar{\Sigma}^{(')0})$                                               &   $4.2\pm 0.6$               & $6.60\pm 0.30$          &    $4.20\pm 0.60$                 &    $4.20\pm 0.60$                                                   &    $2.53\pm 0.28^L$                 &    $2.57\pm 0.35^L$                               \\
$Br(\chi_{c1}\to \Sigma ^{+}\bar{\Sigma}^{-})$                                                    &   $3.6\pm 0.7$               & $6.59\pm 0.30$          &    $3.60\pm 0.70$                 &    $3.60\pm 0.70$                                                  &    $3.12\pm 0.22$                   &    $3.20\pm 0.30$                                 \\
$Br(\chi_{c1}\to \Sigma ^{-}\bar{\Sigma}^{+})$                                                    &   $5.7\pm 1.5$               & $6.59\pm 0.30$          &    $5.70\pm 1.50$                 &    $5.70\pm 1.50$                                                   &    $2.13\pm 0.45^L$                 &    $2.32\pm 0.64^L$                              \\
$Br(\chi_{c1}\to \Xi ^{0}\bar{\Xi}^{0})$                                                          &   $<6$                       & $5.96\pm 0.27$          &    $3.96\pm 2.04$                 &    $3.93\pm 2.07$                                                  &    $3.20\pm 0.28$                   &    $3.27\pm 0.36$                                \\
$Br(\chi_{c1}\to \Xi ^{-}\bar{\Xi}^{+})$                                                          &   $8.0\pm 2.1$               & $5.92\pm 0.27$          &    $8.00\pm 2.09$                 &    $8.00\pm 2.09$                                                  &    $6.18\pm 0.28$                   &    $6.18\pm 0.28$                                \\
$Br(\chi_{c1}\to \Lambda^{(')} \bar{\Sigma}^{(')0}+c.c.)$                                         &   $\cdots$                   & $\cdots$                &    $5.15\pm 4.88$                 &    $4.76\pm 4.36$                                                    &    $0.11\pm 0.02$                   &    $0.16\pm0.03$                                 \\ \hline
$A_{1}$ $(10^{-3}$ GeV$)$                                                                         &   $\cdots$                   & $3.65\pm 0.17$          &    $1.07\pm0.22$                  &    $1.06\pm 0.23$                                                 &    $1.41\pm0.08$                    &    $1.40\pm 0.09$                                \\
$D_{1}$ $(10^{-3}$ GeV$)$                                                                         &   $\cdots$                   & $\cdots$                &    $0.02\pm 2.58$                 &    $0.01\pm 2.50$                                                  &    $0.00\pm 0.29$                   &    $0.00\pm 0.29$                              \\
$F_{1}$ $(10^{-3}$ GeV$)$                                                                         &   $\cdots$                   & $\cdots$                &    $0.01\pm 1.97$                 &    $-0.02\pm 2.00$                                               &    $0.00\pm 0.29$                   &    $0.00\pm 0.29$                               \\
$D_{1}'$ $(10^{-3}$ GeV$)$                                                                        &   $\cdots$                   & $\cdots$                & \begin{tabular}{c}  $-2.12\pm 0.35$\\  or $-0.85\pm 0.14$ \end{tabular}                                 & \begin{tabular}{c}$-2.12\pm 0.34$\\ or $-0.93\pm 0.07$\end{tabular}               &    $-1.96\pm 0.08$                  &    $-1.97\pm0.09$          \\
$F_{1}'$ $(10^{-3}$ GeV$)$                                                                        &   $\cdots$                   & $\cdots$                & \begin{tabular}{c}$0.11\pm1.13$\\  or $3.82\pm0.89$ \end{tabular}                    & \begin{tabular}{c}$-0.34\pm1.13$\\  or $3.93\pm0.78$ \end{tabular}                  &    $0.16\pm0.12$                    &    $0.13\pm0.15$                                \\
$\varphi _{1}$ (radian)                                                                           &   $\cdots$                   & $\cdots$                &    $0\pm\pi$                      &    $0\pm\pi$                   &    $0\pm\pi$                        &    $0\pm\pi$                                         \\
$\alpha$   (radian)                                                                               &   $\cdots$                   & $\cdots$                &    $0$                            &    $0.015\pm0.001$                                                  &    $0$                              &    $0.015\pm0.001$                               \\ \hline
\end{tabular}}\label{Tab:Brchi11S}
%
\caption{The allowed amplitude parameters and branching ratios (in units of $10^{-5}$) of $\protect\chi_{c2}\rightarrow \mathcal{B}_8\bar{\mathcal{B}}_8$ decays within $1\sigma $ errors. }
\centering%
\renewcommand\arraystretch{0.8}\tabcolsep  0.03in
{\scriptsize
\begin{tabular}{lcccc|cc}\hline\hline
              & Exp. data & \begin{tabular}{c} Predictions \\ in $C_1$ case\end{tabular}&\begin{tabular}{c}Predictions  \\ in  $C_2$ case\end{tabular}&\begin{tabular}{c}Predictions  \\ in  $C_3$ case \end{tabular}&\begin{tabular}{c}Predictions  \\ in  $C'_2$ case \end{tabular} &\begin{tabular}{c}Predictions  \\ in  $C'_3$ case \end{tabular}\\ \hline
$Br(\chi_{c2}\to p\bar{p})$                                         &    $7.33\pm 0.33$         &   $7.33\pm 0.33^\dag$       &     $7.33\pm 0.33$             &     $7.33\pm 0.33$                                            &     $7.33\pm 0.33$        &      $7.33\pm 0.33$                                 \\
$Br(\chi_{c2}\to n\bar{n})$                                         &    $\cdots$               &   $7.33\pm 0.33$            &     $5.64\pm 5.57$             &     $5.72\pm 5.49$                                            &     $8.52\pm 3.75$        &      $8.57\pm 3.78$                                \\
$Br(\chi_{c2}\to \Lambda^{(')} \bar{\Lambda}^{(')})$                &    $18.3\pm 1.6$          &   $6.72\pm 0.30$            &     $18.30\pm 1.60$            &     $18.30\pm 1.60$                                         &     $18.30\pm 1.60$       &      $18.30\pm 1.60$                             \\
$Br(\chi_{c2}\to \Sigma^{(')0}\bar{\Sigma}^{(')0})$                 &    $3.7\pm 0.6$           &   $6.40\pm 0.29$            &     $3.70\pm 0.60$             &     $3.70\pm 0.60$                                            &     $3.70\pm 0.60$        &      $3.70\pm 0.60$                               \\
$Br(\chi_{c2}\to \Sigma ^{+}\bar{\Sigma}^{-})$                      &    $3.4\pm 0.7$           &   $6.40\pm 0.29$            &     $3.40\pm 0.70$             &     $3.40\pm 0.70$                                             &     $3.40\pm 0.70$        &      $3.40\pm 0.70$                                \\
$Br(\chi_{c2}\to \Sigma ^{-}\bar{\Sigma}^{+})$                      &    $4.4\pm 1.8$           &   $6.40\pm 0.29$            &     $4.40\pm 1.79$             &     $4.40\pm 1.79$                                            &     $3.87\pm 1.27$        &      $3.86\pm 1.26$                                 \\
$Br(\chi_{c2}\to \Xi ^{0}\bar{\Xi}^{0})$                            &    $<10$                  &   $5.81\pm 0.26$            &     $7.46\pm 2.54$             &     $7.69\pm 2.31$                                            &     $7.43\pm 2.51$        &      $7.31\pm 2.39$                                 \\
$Br(\chi_{c2}\to \Xi ^{-}\bar{\Xi}^{+})$                            &    $14.2\pm 3.2$          &   $5.77\pm 0.26$            &     $14.20\pm 3.20$            &     $14.20\pm3.20$                                            &     $6.84\pm 3.48^L$      &      $6.78\pm 3.53^L$                            \\
$Br(\chi_{c2}\to \Lambda^{(')} \bar{\Sigma}^{(')0}+c.c)$            &    $\cdots$               &   $\cdots$                  &     $2.66\pm 2.59$             &     $2.66\pm 2.56$                                            &     $0.10\pm 0.10$        &      $0.11\pm 0.11$                               \\ \hline
$A_{2}$ $(10^{-3}$  GeV$)$                                          &    $\cdots$               &   $5.51\pm 0.25$            &     $2.53\pm 0.46$             &      $2.53\pm 0.47$                                           &     $2.46\pm 0.42$        &      $2.46\pm 0.41$                               \\
$D_{2}$ $(10^{-3}$  GeV$)$                                          &    $\cdots$               &   $\cdots$                  &     $0.03\pm 2.79$             &      $0.03\pm2.79$                                            &     $0.00\pm 0.56$        &      $0.00\pm 0.56$                              \\
$F_{2}$ $(10^{-3}$   GeV$)$                                         &    $\cdots$               &   $\cdots$                  &     $-0.06\pm 2.99$            &      $-0.06\pm 2.99$                                           &     $0.00\pm 0.55$        &      $0.00\pm 0.55$                                \\
$D_{2}'$ $(10^{-3}$  GeV$)$                                         &   $\cdots$                &   $\cdots$                  &     $-3.56\pm 0.49$            &      $-3.57\pm0.50$                                           &     $-3.23\pm 0.34$       &      $-3.23\pm0.34$                              \\
$F_{2}'$ $(10^{-3}$   GeV$)$                                        &    $\cdots$               &   $\cdots$                  &     $-1.55\pm 1.60$            &      $-1.52\pm1.57$                                           &     $-0.13\pm 0.43$       &      $-0.13\pm0.43$                             \\
$\varphi _{2}$ (radian)                                             &    $\cdots$               &   $\cdots$                  &     $0\pm\pi$                  &      $0\pm\pi$                                                     &     $0\pm\pi$             &      $0\pm\pi$                                         \\
$\alpha$   (radian)                                                 &    $\cdots$               &   $\cdots$                  &     $0$                        &      $0.015\pm0.001$                                          &     $0$                   &      $0.015\pm0.001$                             \\ \hline
\end{tabular} }\label{Tab:Brchi21S}
\end{table}

\clearpage
\section{$\chi _{cJ}\to \mathcal{B}_{10}\bar{\mathcal{B}}_{10}$  decays}

\subsection{Amplitude Relations}
The light baryon decuplet  under the SU(3) flavor symmetry of $u,d,s$ quarks
can be written as
\begin{equation}
\mathcal{B}_{10}=\frac{1}{\sqrt{3}}\left( \left(
\begin{array}{ccc}
\sqrt{3}\Delta ^{++} & \Delta ^{+} & \Sigma^{\ast +} \\
\Delta ^{+} & \Delta ^{0} & \frac{\Sigma^{\ast 0}}{\sqrt{2}} \\
\Sigma ^{\ast +} & \frac{\Sigma ^{\ast 0}}{\sqrt{2}} & \Xi ^{\ast 0}
\end{array}
\right) ,\left(
\begin{array}{ccc}
\Delta ^{+} & \Delta ^{0} & \frac{\Sigma^{\ast 0}}{\sqrt{2}} \\
\Delta ^{0} & \sqrt{3}\Delta ^{-} & \Sigma ^{\ast -} \\
\frac{\Sigma^{\ast 0}}{\sqrt{2}} & \Sigma ^{\ast -} & \Xi ^{\ast -}
\end{array}
\right) ,\left(
\begin{array}{ccc}
\Sigma ^{\ast +} & \frac{\Sigma^{\ast 0}}{\sqrt{2}} & \Xi ^{\ast 0} \\
\frac{\Sigma^{\ast 0}}{\sqrt{2}} & \Sigma ^{\ast -} & \Xi ^{\ast -} \\
\Xi ^{\ast 0} & \Xi ^{\ast -} & \sqrt{3}\Omega ^{-}
\end{array}
\right) \right).   \label{III-1}
\end{equation}%
According to the studies in Refs. \cite{Mo:2023wrf,Mo:2021asa}, the  effective interaction Hamiltonian
for the $\chi _{cJ}\rightarrow \mathcal{B}_{10}\bar{\mathcal{B}}_{10}$ decays can be represented as
\begin{eqnarray}
\mathcal{H}_{eff}\left( \chi _{cJ}\rightarrow \mathcal{B}_{10}\bar{\mathcal{B}}_{10}\right)
=\hat{g}_{0}\mathcal{B}_{10}^{ijk}\bar{\mathcal{B}}_{10}^{ijk}+\hat{g}_{m}H_{3}^{3}+\hat{g}_{e}H_{1}^{1},
\label{III-2}
\end{eqnarray}
with
\begin{eqnarray}
H_{3}^{3}&=&\mathcal{B}_{10}^{3jk}\bar{\mathcal{B}}_{10}^{3jk}-\frac{1}{3}(\mathcal{B}_{10}^{ijk}\bar{\mathcal{B}}
_{10}^{ijk}),  \label{III-3}\\
H_{1}^{1}&=&\mathcal{B}_{10}^{1jk}\bar{\mathcal{B}}_{10}^{1jk}-\frac{1}{3}(\mathcal{B}_{10}^{ijk}\bar{\mathcal{B}}
_{10}^{ijk}),  \label{III-4}
\end{eqnarray}
where $i$ is the number of matrices, $j$ and $k$ are rows and columns,
respectively.  $\hat{g}_{0,m,e}$  are  nonperturbative coupling coefficients for the $\chi _{cJ}\rightarrow \mathcal{B}_{10}\bar{\mathcal{B}}_{10}$ decays.
$\hat{g}_{0}$ is one  under the SU(3) flavor symmetry, and $\hat{g}_{m,e}$ are ones from the SU(3) flavor breaking (mass breaking $\hat{g}_{m}$ and  electromagnetic breaking $\hat{g}_{e}$).
By using Eq. (\ref{III-2}), the amplitudes of the $\chi _{cJ}\rightarrow \mathcal{B}_{10}\bar{\mathcal{B}}_{10}$ decays are obtained, and they  are listed in  Tab. \ref{Tab:AmpT10}.
\begin{table}[h]
\renewcommand\arraystretch{1.1}
\tabcolsep 0.25in
\caption{ Amplitudes  for the $\chi _{cJ}\rightarrow \mathcal{B}_{10}\bar{\mathcal{B}}_{10}$ decays. $\hat{A}_J=\hat{g}_{0}$, $\hat{D}_J=\hat{g}_{e}/3$, and $\hat{D}'_J=\hat{g}_{m}/3$.}
\begin{center}
\begin{tabular}{lc|lc}
\hline\hline
Decay modes                                                             & Amplitudes                                   &   Decay modes                                                    & Amplitudes \\ \hline
$\chi _{cJ}\rightarrow\Delta^{++}     \bar{\Delta}^{--}$                & $\hat{A}_J+2\hat{D}_J-\hat{D}'_J$             &   $\chi _{cJ}\rightarrow\Sigma^{\ast 0}\bar{\Sigma}^{\ast 0}$    & $\hat{A}_J$ \\
$\chi _{cJ}\rightarrow\Delta^{+}      \bar{\Delta}^{-}$                 & $\hat{A}_J+\hat{D}_J-\hat{D}'_J$             &   $\chi _{cJ}\rightarrow\Sigma^{\ast -}\bar{\Sigma}^{\ast +}$    & $\hat{A}_J-\hat{D}_J$ \\
$\chi _{cJ}\rightarrow\Delta^{0}      \bar{\Delta}^{0}$                 & $\hat{A}_J-\hat{D}'_J$                       &   $\chi _{cJ}\rightarrow\Xi^{\ast 0}\bar{\Xi}^{\ast 0}$          & $\hat{A}_J+\hat{D}'_J$ \\
$\chi _{cJ}\rightarrow\Delta^{-}      \bar{\Delta}^{+}$                 & $\hat{A}_J-\hat{D}_J-\hat{D}'_J$             &   $\chi _{cJ}\rightarrow\Xi^{\ast -}\bar{\Xi}^{\ast +}$          & $\hat{A}_J-\hat{D}_J+\hat{D}'_J$ \\
$\chi _{cJ}\rightarrow\Sigma^{\ast +} \bar{\Sigma}^{\ast -}$            & $\hat{A}_J+\hat{D}_J$                        &   $\chi _{cJ}\rightarrow\Omega^{-}\bar{\Omega}^{+}$              & $\hat{A}_J-\hat{D}_J+2\hat{D}'_J$ \\ \hline
\end{tabular}
\end{center}\label{Tab:AmpT10}
\end{table}

\subsection{Numerical result and discussion}

The theoretical input parameters and the
experimental data within $2\sigma$ errors and  $1\sigma$ error from PDG \cite{ParticleDataGroup:2022pth}
will be used for the  numerical analysis of $\chi _{cJ}\rightarrow \mathcal{B}_{10}\bar{\mathcal{B}}_{10}$ decays.

As for the $\chi _{cJ}\rightarrow \mathcal{B}_{10}\bar{\mathcal{B}}_{10}$ decays, there are only four free parameters $\left( \hat{A}_{J},\hat{D}_{J},\hat{D}_{J}',\hat{\varphi} _{J}\right)$.
Three measured decay modes exist in the $\chi _{c0}\rightarrow \mathcal{B}_{10}\bar{\mathcal{B}}_{10}$ decays, along with one experimental data point and two experimental upper limits in the $\chi _{c1}\rightarrow \mathcal{B}_{10}\bar{\mathcal{B}}_{10}$  or $\chi _{c2}\rightarrow \mathcal{B}_{10}\bar{\mathcal{B}}_{10}$ decays.
$Br(\chi _{cJ}\rightarrow \Sigma ^{\ast \pm }\bar{\Sigma}^{\ast \mp })$ from PDG \cite{ParticleDataGroup:2022pth} and  $Br(\chi _{cJ}\rightarrow \Omega ^{-}\bar{\Omega}^{+})$  from the BESIII Collaboration \cite{BESIII:2023olq} will be used for our numerical analysis.
The experimental data within  $2\sigma$ errors ($1\sigma$ error) are listed in the second columns of Tab. \ref{Tab:Brchi0102S}, Tab. \ref{Tab:Brchi1102S}, and Tab. \ref{Tab:Brchi2102S} (Tab. \ref{Tab:Brchi0101S}, Tab. \ref{Tab:Brchi1101S}, and Tab. \ref{Tab:Brchi2101S}).

\begin{table}[p]
\renewcommand\arraystretch{1.0}
\tabcolsep 0.05in
\caption{The allowed amplitude parameters and branching ratios (in units of $10^{-4}$) of the $\chi_{c0}\rightarrow \mathcal{B}_{10}\bar{\mathcal{B}}_{10}$ decays within $2\sigma $ errors. }
\begin{center}
\begin{tabular}{lccc|c}
\hline\hline
                                                                 & Exp. data               & Predictions in $S_1$ case    & Predictions in $S_2$ case     & Predictions in $S'_2$ case    \\ \hline
$Br(\chi _{c0}\to \Delta ^{++}\bar{\Delta}^{--})$                & $\cdots$                & $1.35\pm 0.29$               & $5.69\pm 5.51$                & $4.34\pm 4.16$              \\
$Br(\chi _{c0}\to \Delta ^{+}\bar{\Delta}^{-})$                  & $\cdots$                & $1.35\pm 0.29$               & $5.85\pm 5.85$                & $3.57\pm 3.20$              \\
$Br(\chi _{c0}\to \Delta ^{0}\bar{\Delta}^{0})$                  & $\cdots$                & $1.35\pm 0.29$               & $6.33\pm 6.33$                & $3.63\pm 3.10$              \\
$Br(\chi _{c0}\to \Delta ^{-}\bar{\Delta}^{+})$                  & $\cdots$                & $1.35\pm 0.29$               & $7.37\pm 6.52$                & $4.59\pm 3.75$              \\
$Br(\chi _{c0}\to \Sigma ^{\ast +}\bar{\Sigma}^{\ast -})$        & $1.60\pm 1.20$          & $1.14\pm 0.24$               & $1.60\pm 1.20$                & $1.60\pm 1.20$              \\
$Br(\chi _{c0}\to \Sigma ^{\ast 0}\bar{\Sigma}^{\ast 0})$        & $\cdots$                & $1.15\pm 0.24$               & $1.60\pm 1.60$                & $1.87\pm 1.35$              \\
$Br(\chi _{c0}\to \Sigma ^{\ast -}\bar{\Sigma}^{\ast +})$        & $2.30\pm 1.40$          & $1.14\pm 0.24$               & $2.30\pm 1.40$                & $2.30\pm 1.40$              \\
$Br(\chi _{c0}\to \Xi ^{\ast 0}\bar{\Xi}^{\ast 0})$              & $\cdots$                & $0.86\pm 0.18$               & $0.75\pm 0.75$                & $0.92\pm 0.59$              \\
$Br(\chi _{c0}\to \Xi ^{\ast -}\bar{\Xi}^{\ast +})$              & $\cdots$                & $0.86\pm 0.18$               & $0.88\pm 0.88$                & $1.19\pm 0.61$              \\
$Br(\chi _{c0}\to \Omega ^{-}\bar{\Omega}^{+})$                  & $0.351\pm 0.122$        & $0.390\pm 0.083$             & $0.351\pm 1.22$               & $0.351\pm 0.122$            \\     \hline
$\hat{A}_{0}$ $(10^{-2}$ GeV$)$                                  & $\cdots$                & $1.90\pm 0.30$               & $1.70\pm 1.60$                & $2.29\pm 1.03$              \\
$\hat{D}_{0}$ $(10^{-2}$ GeV$)$                                  & $\cdots$                & $\cdots$                     & $0.00\pm 2.99$                & $0.02\pm 1.57$              \\
$\hat{D}'_{0}$ $(10^{-2}$ GeV$)$                                 & $\cdots$                & $\cdots$                     & $-0.23\pm 2.58$               & $-0.66\pm 0.86$              \\    \hline
\end{tabular}\label{Tab:Brchi0102S}
\end{center}
%
\renewcommand\arraystretch{1.0}
\tabcolsep 0.05in
\caption{The allowed amplitude parameters and branching ratios (in units of $10^{-5}$) of the $\chi_{c1}\rightarrow \mathcal{B}_{10}\bar{\mathcal{B}}_{10}$ decays within $2\sigma $ errors. }
\begin{center}
\begin{tabular}{lccc|c}
\hline\hline
                                                                  & Exp. data               & Predictions in $S_1$ case    & Predictions in $S_2$ case         & Predictions in $S'_2$ case         \\ \hline
$Br(\chi _{c1}\to \Delta ^{++}\bar{\Delta}^{--} )$                & $\cdots$                & $3.50\pm 1.19$               & $<34.99$                          & $<20.76$                             \\
$Br(\chi _{c1}\to \Delta ^{+}\bar{\Delta}^{-})$                   & $\cdots$                & $3.50\pm 1.19$               & $<29.47$                          & $<13.04$                             \\
$Br(\chi _{c1}\to \Delta ^{0}\bar{\Delta}^{0} )$                  & $\cdots$                & $3.50\pm 1.19$               & $<24.45$                          & $<10.15$                             \\
$Br(\chi _{c1}\to \Delta ^{-}\bar{\Delta}^{+} )$                  & $\cdots$                & $3.50\pm 1.19$               & $<21.61$                          & $<8.32$                             \\
$Br(\chi _{c1}\to \Sigma ^{\ast +}\bar{\Sigma}^{\ast -} )$        & $<9.00$                 & $3.02\pm 1.02$               & $<9.00$                           & $<9.00$                             \\
$Br(\chi _{c1}\to \Sigma ^{\ast 0}\bar{\Sigma}^{\ast 0} )$        & $\cdots$                & $3.02\pm 1.02$               & $<6.74$                           & $<6.74$                             \\
$Br(\chi _{c1}\to \Sigma ^{\ast -}\bar{\Sigma}^{\ast +} )$        & $<5.00$                  & $3.02\pm 1.02$               & $<5.00$                          & $<5.00$                             \\
$Br(\chi _{c1}\to \Xi ^{\ast 0}\bar{\Xi}^{\ast 0} )$              & $\cdots$                & $2.40\pm 0.81$               & $<4.84$                           & $<4.92$                             \\
$Br(\chi _{c1}\to \Xi ^{\ast -}\bar{\Xi}^{\ast +} )$              & $\cdots$                & $2.38\pm 0.81$               & $<3.52$                           & $<3.53$                             \\
$Br(\chi _{c1}\to \Omega ^{-}\bar{\Omega}^{+} )$                  & $1.49\pm 0.50$          & $1.49\pm 0.50$               & $1.49\pm 0.50$                    & $1.49\pm 0.50$                    \\ \hline
$\hat{A}_{1}$ $(10^{-3}$ GeV$)$                                   & $\cdots$                & $2.68\pm 0.59$               & $2.11\pm 2.11$                    & $2.51\pm 1.65$                     \\
$\hat{D}_{1}$ $(10^{-3}$ GeV$)$                                   & $\cdots$                & $\cdots$                     & $0.02\pm 4.17$                    & $0.02\pm 1.93$                     \\
$\hat{D}'_{1} (10^{-3}$ GeV$)$                                    & $\cdots$                & $\cdots$                     & $-0.10\pm 3.29$                   & $0.15\pm 0.94$                     \\ \hline
\end{tabular}\label{Tab:Brchi1102S}
\end{center}
%
\renewcommand\arraystretch{1.0}
\tabcolsep 0.05in
\caption{The allowed amplitude parameters and branching ratios (in units of $10^{-5}$) of the $\chi_{c2}\rightarrow \mathcal{B}_{10}\bar{\mathcal{B}}_{10}$ decays within $2\sigma $ errors. }
\begin{center}
\begin{tabular}{lccc|c}
\hline\hline
                                                                  & Exp. data               & Predictions in $S_1$ case             & Predictions in $S_2$ case         & Predictions in $S'_2$ case        \\ \hline
$Br(\chi _{c2}\to \Delta ^{++}\bar{\Delta}^{--} )$                & $\cdots$                & $8.76\pm 0.46$                        & $<65.04$                          & $<29.79$                          \\ \hline
$Br(\chi _{c2}\to \Delta ^{+}\bar{\Delta}^{-} )$                  & $\cdots$                & $8.76\pm 0.46$                        & $<52.23$                          & $<18.07$                          \\ \hline
$Br(\chi _{c2}\to \Delta ^{0}\bar{\Delta}^{0} )$                  & $\cdots$                & $8.76\pm 0.46$                        & $<43.39$                          & $<13.07$                          \\ \hline
$Br(\chi _{c2}\to \Delta ^{-}\bar{\Delta}^{+} )$                  & $\cdots$                & $8.76\pm 0.46$                        & $<37.90$                          & $<9.58$                          \\ \hline
$Br(\chi _{c2}\to \Sigma ^{\ast +}\bar{\Sigma}^{\ast -} )$        & $<16.00$                & $7.61\pm 0.39$                        & $<16.00$                          & $<16.00$                          \\ \hline
$Br(\chi _{c2}\to \Sigma ^{\ast 0}\bar{\Sigma}^{\ast 0} )$        & $\cdots$                & $7.63\pm 0.39$                        & $<11.18$                          & $<11.23$                          \\ \hline
$Br(\chi _{c2}\to \Sigma ^{\ast -}\bar{\Sigma}^{\ast +} )$        & $<8.00$                 & $7.61\pm 0.39$                        & $<8.00$                           & $<8.00$                          \\ \hline
$Br(\chi _{c2}\to \Xi ^{\ast 0}\bar{\Xi}^{\ast 0} )$              & $\cdots$                & $6.17\pm 0.32$                        & $<9.85$                           & $<9.95$                          \\ \hline
$Br(\chi _{c2}\to \Xi ^{\ast -}\bar{\Xi}^{\ast +} )$              & $\cdots$                & $6.13\pm 0.32$                        & $<6.99$                           & $<7.01$                          \\ \hline
$Br(\chi _{c2}\to \Omega ^{-}\bar{\Omega}^{+} )$                  & $4.52\pm 0.60$          & $4.13\pm 0.21$                        & $4.52\pm0.60$                     & $4.52\pm 0.60$                  \\ \hline
$A_{2}$ $(10^{-3}$ GeV$)$                                         & $\cdots$                & $6.54\pm 0.46$                        & $4.09\pm 4.09$                    & $5.35\pm 2.78$                        \\ \hline
$D_{2}$ $(10^{-3}$ GeV$)$                                         & $\cdots$               & $\cdots$                               & $0.02\pm 8.17$                    & $0.03\pm  3.72$               \\ \hline
$D'_{2}$ $(10^{-3}$ GeV$)$                                        & $\cdots$                & $\cdots$                              & $-0.05\pm7.10$                    & $1.18\pm  1.32$                  \\ \hline
\end{tabular}\label{Tab:Brchi2102S}
\end{center}
\end{table}

\begin{table}[htbp]
\renewcommand\arraystretch{1.1}
\tabcolsep 0.1in
\caption{The allowed amplitude parameters and branching ratios (in units of $10^{-4}$) of the $\chi_{c0}\rightarrow \mathcal{B}_{10}\bar{\mathcal{B}}_{10}$ decays within $1\sigma $ error. $^\sharp$Its experimental data are not used in the $S_1$ case. }
\begin{center}
\begin{tabular}{lccc|c}
\hline\hline
                                                                 & Exp. data               & Predictions in $S_1$ case    & Predictions in $S_2$ case     & Predictions in $S'_2$ case    \\ \hline
$Br(\chi _{c0}\to \Delta ^{++}\bar{\Delta}^{--})$                & $\cdots$                & $1.21\pm 0.22$               & $5.06\pm 3.96$                & $3.43\pm 2.34$              \\
$Br(\chi _{c0}\to \Delta ^{+}\bar{\Delta}^{-})$                  & $\cdots$                & $1.21\pm 0.22$               & $4.79\pm 4.79$                & $2.87\pm 1.45$              \\
$Br(\chi _{c0}\to \Delta ^{0}\bar{\Delta}^{0})$                  & $\cdots$                & $1.21\pm 0.22$               & $5.22\pm 5.21$                & $3.11\pm 1.59$              \\
$Br(\chi _{c0}\to \Delta ^{-}\bar{\Delta}^{+})$                  & $\cdots$                & $1.21\pm 0.22$               & $6.89\pm 4.72$                & $3.92\pm 1.75$              \\
$Br(\chi _{c0}\to \Sigma ^{\ast +}\bar{\Sigma}^{\ast -})$        & $1.60\pm 0.60$          & $1.02\pm 0.18^\sharp$        & $1.60\pm 0.60$                & $1.60\pm 0.60$              \\
$Br(\chi _{c0}\to \Sigma ^{\ast 0}\bar{\Sigma}^{\ast 0})$        & $\cdots$                & $1.02\pm 0.18$               & $1.27\pm 1.27$                & $1.80\pm 0.78$              \\
$Br(\chi _{c0}\to \Sigma ^{\ast -}\bar{\Sigma}^{\ast +})$        & $2.30\pm 0.70$          & $1.02\pm 0.18^\sharp$        & $2.30\pm 0.70$                & $2.30\pm 0.70$              \\
$Br(\chi _{c0}\to \Xi ^{\ast 0}\bar{\Xi}^{\ast 0})$              & $\cdots$                & $0.77\pm 0.14$               & $0.62\pm 0.62$                & $0.87\pm 0.38$              \\
$Br(\chi _{c0}\to \Xi ^{\ast -}\bar{\Xi}^{\ast +})$              & $\cdots$                & $0.77\pm 0.14$               & $0.75\pm 0.74$                & $1.19\pm 0.30$              \\
$Br(\chi _{c0}\to \Omega ^{-}\bar{\Omega}^{+})$                  & $0.351\pm 0.061$        & $0.351\pm 0.061$             & $0.351\pm 0.061$              & $0.351\pm 0.061$            \\     \hline
$\hat{A}_{0}$ $(10^{-2}$ GeV$)$                                  & $\cdots$                & $1.80\pm 0.21$               & $1.50\pm 1.40$                & $2.35\pm 0.58$              \\
$\hat{D}_{0}$ $(10^{-2}$ GeV$)$                                  & $\cdots$                & $\cdots$                     & $0.00\pm 2.90$                & $0.00\pm 1.37$              \\
$\hat{D}'_{0}$ $(10^{-2}$ GeV$)$                                 & $\cdots$                & $\cdots$                     & $-0.23\pm 2.30$               & $-0.53\pm 0.37$              \\    \hline
\end{tabular}\label{Tab:Brchi0101S}
\end{center}
%
\renewcommand\arraystretch{1.1}
\tabcolsep 0.1in
\caption{The allowed amplitude parameters and branching ratios (in units of $10^{-5}$) of the $\chi_{c1}\rightarrow \mathcal{B}_{10}\bar{\mathcal{B}}_{10}$ decays within $1\sigma $ error. }
\begin{center}
\begin{tabular}{lccc|c}
\hline\hline
                                                                  & Exp. data               & Predictions in $S_1$ case    & Predictions in $S_2$ case         & Predictions in $S'_2$ case         \\ \hline
$Br(\chi _{c1}\to \Delta ^{++}\bar{\Delta}^{--} )$                & $\cdots$                & $3.50\pm 0.60$               & $<33.88$                          & $<20.46$                             \\
$Br(\chi _{c1}\to \Delta ^{+}\bar{\Delta}^{-})$                   & $\cdots$                & $3.50\pm 0.60$               & $<28.26$                          & $<12.85$                             \\
$Br(\chi _{c1}\to \Delta ^{0}\bar{\Delta}^{0} )$                  & $\cdots$                & $3.50\pm 0.60$               & $<24.22$                          & $<9.85$                             \\
$Br(\chi _{c1}\to \Delta ^{-}\bar{\Delta}^{+} )$                  & $\cdots$                & $3.50\pm 0.60$               & $<21.07$                          & $<7.74$                             \\
$Br(\chi _{c1}\to \Sigma ^{\ast +}\bar{\Sigma}^{\ast -} )$        & $<9.00$                 & $3.01\pm 0.51$               & $<9.00$                           & $<9.00$                             \\
$Br(\chi _{c1}\to \Sigma ^{\ast 0}\bar{\Sigma}^{\ast 0} )$        & $\cdots$                & $3.02\pm 0.51$               & $<6.65$                           & $<6.79$                             \\
$Br(\chi _{c1}\to \Sigma ^{\ast -}\bar{\Sigma}^{\ast +} )$        & $<5.00$                  & $3.01\pm 0.51$               & $<5.00$                          & $<5.00$                             \\
$Br(\chi _{c1}\to \Xi ^{\ast 0}\bar{\Xi}^{\ast 0} )$              & $\cdots$                & $2.40\pm 0.41$               & $<4.68$                           & $<4.68$                             \\
$Br(\chi _{c1}\to \Xi ^{\ast -}\bar{\Xi}^{\ast +} )$              & $\cdots$                & $2.38\pm 0.40$               & $<3.32$                           & $<3.34$                             \\
$Br(\chi _{c1}\to \Omega ^{-}\bar{\Omega}^{+} )$                  & $1.49\pm 0.25$          & $1.49\pm 0.50$               & $1.49\pm 0.25$                    & $1.49\pm 0.25$                    \\ \hline
$\hat{A}_{1}$ $(10^{-3}$ GeV$)$                                   & $\cdots$                & $2.69\pm 0.29$               & $2.03\pm 2.03$                    & $2.55\pm 1.57$                     \\
$\hat{D}_{1}$ $(10^{-3}$ GeV$)$                                   & $\cdots$                & $\cdots$                     & $-0.01\pm 4.04$                    & $0.00\pm 1.95$                     \\
$\hat{D}'_{1} (10^{-3}$ GeV$)$                                    & $\cdots$                & $\cdots$                     & $-0.04\pm 3.16$                   & $0.21\pm 0.82$                     \\ \hline
\end{tabular}\label{Tab:Brchi1101S}
\end{center}
%
\renewcommand\arraystretch{1.1}
\tabcolsep 0.1in
\caption{The allowed amplitude parameters and branching ratios (in units of $10^{-5}$) of the $\chi_{c2}\rightarrow \mathcal{B}_{10}\bar{\mathcal{B}}_{10}$ decays within $1\sigma $ error. }
\begin{center}
\begin{tabular}{lccc|c}
\hline\hline
                                                                  & Exp. data               & Predictions in $S_1$ case             & Predictions in $S_2$ case         & Predictions in $S'_2$ case        \\ \hline
$Br(\chi _{c2}\to \Delta ^{++}\bar{\Delta}^{--} )$                & $\cdots$                & $9.08\pm 0.14$                        & $<64.30$                          & $<32.01$                          \\ \hline
$Br(\chi _{c2}\to \Delta ^{+}\bar{\Delta}^{-} )$                  & $\cdots$                & $9.08\pm 0.14$                        & $<51.89$                          & $<18.40$                          \\ \hline
$Br(\chi _{c2}\to \Delta ^{0}\bar{\Delta}^{0} )$                  & $\cdots$                & $9.08\pm 0.14$                        & $<43.71$                          & $<13.44$                          \\ \hline
$Br(\chi _{c2}\to \Delta ^{-}\bar{\Delta}^{+} )$                  & $\cdots$                & $9.08\pm 0.14$                        & $<37.82$                          & $<9.32$                          \\ \hline
$Br(\chi _{c2}\to \Sigma ^{\ast +}\bar{\Sigma}^{\ast -} )$        & $<16.00$                & $7.89\pm 0.11$                        & $<16.00$                          & $<16.00$                          \\ \hline
$Br(\chi _{c2}\to \Sigma ^{\ast 0}\bar{\Sigma}^{\ast 0} )$        & $\cdots$                & $7.90\pm 0.11$                        & $<10.91$                          & $<11.59$                          \\ \hline
$Br(\chi _{c2}\to \Sigma ^{\ast -}\bar{\Sigma}^{\ast +} )$        & $<8.00$                 & $7.89\pm 0.11$                        & $<8.00$                           & $<8.00$                          \\ \hline
$Br(\chi _{c2}\to \Xi ^{\ast 0}\bar{\Xi}^{\ast 0} )$              & $\cdots$                & $6.39\pm 0.09$                        & $<9.73$                           & $<9.79$                          \\ \hline
$Br(\chi _{c2}\to \Xi ^{\ast -}\bar{\Xi}^{\ast +} )$              & $\cdots$                & $6.35\pm 0.09$                        & $<6.78$                           & $<6.78$                          \\ \hline
$Br(\chi _{c2}\to \Omega ^{-}\bar{\Omega}^{+} )$                  & $4.52\pm 0.30$          & $4.28\pm 0.06$                        & $4.52\pm0.30$                     & $4.52\pm 0.30$                  \\ \hline
$A_{2}$ $(10^{-3}$ GeV$)$                                         & $\cdots$                & $6.66\pm 0.19$                        & $3.95\pm 3.95$                    & $5.32\pm 2.83$                        \\ \hline
$D_{2}$ $(10^{-3}$ GeV$)$                                         & $\cdots$               & $\cdots$                               & $-0.03\pm 7.85$                    & $0.08\pm  3.87$               \\ \hline
$D'_{2}$ $(10^{-3}$ GeV$)$                                        & $\cdots$                & $\cdots$                              & $-0.24\pm6.74$                    & $1.22\pm  1.26$                  \\ \hline
\end{tabular}\label{Tab:Brchi2101S}
\end{center}
\end{table}

The method of obtaining numerical results is similar to ones  of the $\chi_{cJ}\rightarrow \mathcal{B}_{8}\bar{\mathcal{B}}_{8}$ decays.
Two cases are considered  for the $\chi _{cJ}\rightarrow \mathcal{B}_{10}\bar{\mathcal{B}}_{10}$ numerical results. {\bf In $S_1$ case},  only the SU(3) flavor symmetry contribution are considered,  $i.e.$, the coupling  coefficients $\hat{A}_J$ are free parameters, and $\hat{D}_J=\hat{D}'_J=0$.  {\bf In $S_2$ case},  both the SU(3) flavor symmetry and the SU(3) flavor breaking contributions are considered. We set $\hat{A}_J>=0$ and other three parameters are free parameters.

The results  within $2\sigma$ errors in $S_1$ and $S_2$ cases are listed in the third and forth columns of Tab. \ref{Tab:Brchi0102S}, Tab. \ref{Tab:Brchi1102S}, and Tab. \ref{Tab:Brchi2102S}.
From the results in the $S_2$ case, it can be seen that the channels of $\chi _{c0}$ have significantly different branching ratios, which is attributed to the sufficient ability of modes $\Sigma ^{\ast +}\bar{\Sigma}%
^{\ast -}$, $\Sigma ^{\ast -}\bar{\Sigma}^{\ast +}$, and $\Omega ^{-}\bar{\Omega}^{+}$ to reflect the contributions of various amplitude parameters. But the same situation did not occur in the $\chi _{c1}$ and $\chi _{c2}$ decays, the
branching ratios of majority of decay channels that we can only provide upper limits, because there are only the experimental upper limits of $Br(\chi _{c2,3}\rightarrow
\Sigma ^{\ast \pm }\bar{\Sigma}^{\ast \mp })$.
Despite this, the results demonstrate the potential of SU(3) flavor analysis in investigating $\chi _{cJ}\rightarrow B_{10}\bar{B}_{10}$, allowing for the extraction of information from limited data through correlations between modes.
Notably, under consideration of theoretical input parameters and experimental data within $2\sigma$ errors, all relevant experimental data within $2\sigma$ errors can be accounted for in both $S_1$ and $S_2$ cases. However, the $S_2$ case exhibits significantly large errors in branching ratio predictions. The phase angles $\hat{\varphi} _{J}$ remain unconstrained by current experimental data and are thus omitted from the tables.

From the results in  the  $S_2$ case, one can see that some absolute values of the constrained  $\hat{D}_{J},\hat{D}_{J}'$ are not far less than that for $\hat{A}_{J}$.
Referring to charmonium $J/\psi,\Psi(2S)\rightarrow \mathcal{B}_{10}\bar{\mathcal{B}}_{10}$ decays in Ref. \cite{Mo:2023wrf}, the maximum value of   the constrained  $|\hat{D}/\hat{A}|$ and $|\hat{D}'/\hat{A}|$ is 52.30\%. Next, we  assume that the ratio $|\hat{D}_J/\hat{A}_J|,|\hat{D}'_J/\hat{A}_J|$ are less than or equal to 50\% to obtain our numerical results, which is  renamed $S'_{2}$ case.
The results within  $2\sigma$ errors in $S'_2$ case are listed in the last columns of Tab. \ref{Tab:Brchi0102S}-\ref{Tab:Brchi2102S}.
One can see that  the  predictions is more accurate than ones in the $S_2$ case, nevertheless, the errors are still large due to few experimental measurements.

In addition, an analysis of the results within $1\sigma$ error range has been conducted and is documented in Tab. \ref{Tab:Brchi0101S}, Tab. \ref{Tab:Brchi1101S}, and Tab. \ref{Tab:Brchi2101S} for the $\chi_{c0}\rightarrow \mathcal{B}_{10}\bar{\mathcal{B}}_{10}$, $\chi_{c1}\rightarrow \mathcal{B}_{10}\bar{\mathcal{B}}_{10}$ and $\chi_{c2}\rightarrow \mathcal{B}_{10}\bar{\mathcal{B}}_{10}$ decays, respectively.  For the  $\chi_{c0}\rightarrow \mathcal{B}_{10}\bar{\mathcal{B}}_{10}$ decays in the $S_1$ case, three experimental branching ratios can not be explained at the same time if only considering the SU(3) flavor symmetry effects, so we only use $Br(\chi_{c0}\rightarrow \Omega^-\bar{\Omega}^+)$ to constrain on $\hat{A}_0$ and to predict other branching ratios, and  one can see that the predicted $Br(\chi _{c0}\to \Sigma ^{\ast +}\bar{\Sigma}^{\ast -})$ lies in its experimental data within $1\sigma$ error, nevertheless, the predicted $Br(\chi _{c0}\to \Sigma ^{\ast -}\bar{\Sigma}^{\ast +})$ exceed  its experimental data within $1\sigma$ error. Comparing these predictions with those within $2\sigma$ error range, the predictions within $1\sigma$ error range exhibit higher accuracy, yet their errors remain substantial. It is anticipated that further measurements of these decays will facilitate more precise predictions.


\section{Conclusion}

The study of charmonium decays into baryon antibaryon  pairs provides a powerful
tool for investigating many topics in quantum chromodynamics.  We have performed an analysis of   the $\chi _{c0,1,2}\to \mathcal{B}_{8}\bar{\mathcal{B}}_{8}$ and $\mathcal{B}_{10}\bar{\mathcal{B}}_{10}$ decays by using  SU(3) flavor analysis, which has also been employed for the $J/\psi $ and $\Psi (2S)$ decays \cite{Mo:2023wrf,Ferroli:2020mra}. We firstly constrain on the relevant coupling coefficients, and then give the predictions of not-measured or not-well-measured branching ratios. The results within both $2\sigma$ errors and $1\sigma$ error are analyzed in different cases.

{\bf For the $\chi _{c0,1,2}\to \mathcal{B}_{8}\bar{\mathcal{B}}_{8}$ decays}, many decay modes are measured at present. However, we discovered that the SU(3) flavor symmetry effects alone cannot explain all experimental data simultaneously. By incorporating SU(3) flavor breaking effects, we successfully account for all experimental data within $2\sigma$ error bounds. Notably, within a $1\sigma$ error range, $Br(\chi_{c0}\to p\bar{p})$ remains unexplained alongside other data. We also investigated $\Sigma^{0}-\Lambda$ mixing with a specified mixing angle $\alpha= (0.015\pm0.001)$ radians. In addition, assuming the breaking coupling $(D_J,F_J,D'_J,F'_J)$ are weaker than the SU(3) symmetry coupling $A_J$, $i.e.$,  $|D_J/A_J|\leq20\%,|F_J/A_J|\leq20\%,|D'_J/A_J|\leq20\%$ and $|F'_J/A_J|\leq20\%$, the results have been reanalyzed.  And we have found that all experimental data can be explained within $2\sigma$ errors, nevertheless, the experimental upper limit of  $Br(\chi_{c0}\to p\bar{p})$ and the lower limits of $Br(\chi_{c1}\to \Sigma^{(')0}\bar{\Sigma}^{(')0})$, $Br(\chi_{c1}\to \Sigma ^{-}\bar{\Sigma}^{+})$  and $Br(\chi_{c1}\to \Sigma ^{-}\bar{\Sigma}^{+})$ can not be explained with other data together within $1\sigma$ error.

{\bf For the $\chi _{c0,1,2}\to \mathcal{B}_{10}\bar{\mathcal{B}}_{10}$ decays}, the experiential data are less, and we only can give the rough results.  All experimental data can be explained in both the SU(3) symmetry and breaking cases within $2\sigma$ errors. Within $1\sigma$ error, $Br(\chi _{c0}\to \Sigma ^{\ast -}\bar{\Sigma}^{\ast +})$  can not explain with other data together under the SU(3) flavor symmetry.  The predictions of  $Br(\chi _{c0,1,2}\to \mathcal{B}_{10}\bar{\mathcal{B}}_{10})$, which have not been measured or well measured, have been given.

Mesons $\chi_{cJ}$  cannot be produced directly in $e^+e^-$ annihilation,  nevertheless, they can be produced in the radiative
decays $\psi(3686)\to \chi_{cJ}\gamma$ with the large branching ratios.  A large sample of $\psi(3686)$ decays at BESIII
 provides very clean environments for the investigation of the
decay mechanism of the $ \chi_{cJ}$ mesons.   According to our predictions, many decay modes  might be measured by the forthcoming BESIII data in the near future.   Additional experimental information will be helpful to reduce the errors of the coupling coefficients and the branching ratio predictions, and then to better understand the decay dynamics of the charmonium decays into baryon and anti-baryon pairs.

\section*{ACKNOWLEDGEMENTS}
We thank Prof. Shuang-Shi Fang for useful discussions. The work was supported by the National Natural Science Foundation of China (No. 12175088 and No. 12365014).

\section*{References}

\end{document}